\documentclass[a4paper,11pt]{article}
\pdfoutput=1 

\usepackage{jcappub} 

\usepackage[T1]{fontenc} 
\usepackage{graphicx}
\graphicspath{{figures/}}
\usepackage[utf8]{inputenc}
\usepackage{amsmath}
\usepackage{amsfonts}
\usepackage{amssymb}
\usepackage{enumitem}
\usepackage{multirow}
\usepackage{CJKutf8}
\usepackage{color}
\usepackage[capitalise]{cleveref}
\usepackage{acro}
\usepackage{adjustbox}
\usepackage{array}
\usepackage{natbib}
\usepackage{dblfnote}
\DFNalwaysdouble
\usepackage{slashed}
\usepackage{dcolumn}
\usepackage{hyperref}
\usepackage{geometry}
\usepackage{subfig} 
\usepackage{overpic}
\usepackage{inputenc}
\usepackage{caption}
\def\({\left(}
\def\){\right)}
\def\[{\left[}
\def\]{\right]}
\def\be{\begin{eqnarray}}
\def\ee{\end{eqnarray}}

\DeclareAcronym{GW}{
  short = GW ,
  long = gravitational wave ,
  short-plural = s 
}
\DeclareAcronym{LIGO}{
  short = LIGO ,
  long = Laser Interferometer Gravitational-wave Observatory ,
  short-plural = 
}

\DeclareAcronym{BBN}{
  short = BBN ,
  long = Big Bang Nucleosynthesis  ,
  short-plural = 
}

\DeclareAcronym{LISA}{
  short = LISA ,
  long = Laser Interferometer Space Antenna ,
  short-plural =  
}
\DeclareAcronym{SKA}{
  short = SKA ,
  long = Square Kilometre Array ,
  short-plural =  
}  
  
\DeclareAcronym{SNR}{
	short = SNR ,
	long = signal-to-noise ratio ,
	short-plural = 
}

\DeclareAcronym{PTA}{
	short = PTA ,
	long = pulsar timing array ,
	short-plural = 
}

\DeclareAcronym{FLRW}{
  short = FLRW ,
  long = Friedmann-Lemaitre-Robertson-Walker ,
  short-plural =  
}

\DeclareAcronym{SIGW}{
	short = SIGW ,
	long = scalar induced gravitational wave ,
	short-plural =  s
}

\DeclareAcronym{PBH}{
	short = PBH ,
	long = primordial black hole ,
	short-plural =  s
}

\DeclareAcronym{CMB}{
	short = CMB ,
	long = cosmic microwave background ,
	short-plural =  
}
\DeclareAcronym{DM}{
	short = DM ,
	long = dark matter ,
	short-plural =  
}

\DeclareAcronym{SGWB}{
	short = SGWB ,
	long = stochastic gravitational	wave background ,
	short-plural =  s
}

\DeclareAcronym{LSS}{
	short = LSS ,
	long = large-scale structure ,
	short-plural =  
}

\DeclareAcronym{RD}{
	short = RD ,
	long = radiation-dominated ,
	short-plural =  
}

\DeclareAcronym{SMBHB}{
  short = SMBHB ,
  long = supermassive black hole binary ,
}

\title{\boldmath Probing scalar induced gravitational waves with PTA and LISA: The Importance of third order correction}

\author[a,b]{Zhe Chang,}
\author[a,b,1]{Yu-Ting Kuang,\note{Corresponding author.}}
\author[a,b]{Di Wu,}
\author[a,b]{Jing-Zhi Zhou}

\affiliation[a]{Institute of High Energy Physics, Chinese Academy of Sciences, Beijing 100049, China}
\affiliation[b]{University of Chinese Academy of Sciences, Beijing 100049, China}

\emailAdd{changz@ihep.ac.cn}
\emailAdd{kuangyt@ihep.ac.cn}
\emailAdd{wudi@ihep.ac.cn}
\emailAdd{zhoujingzhi@ihep.ac.cn}

\abstract{We revisit the calculation of third order \acp{SIGW} and extend it from a monochromatic primordial power spectrum to a more general log-normal one. We investigate the impact of third order SIGWs on \ac{SNR} of \ac{LISA}  and \ac{PTA} observations, and find that third order SIGWs significantly contribute to the total energy density spectrum of \acp{GW} in high-frequency region. For a primordial power spectrum amplitude of $A_{\zeta}=10^{-2}\sim 10^{-1}$, the effects of third order SIGWs lead to a $40\%$ to $400\%$ increase in the SNR for LISA. Additionally, our PTA data analysis reveals that third order SIGWs diminish both the amplitude $A_{\zeta}$ and the peak frequency $f_*$ of the primordial power spectrum.}

\begin{document}
\maketitle
\flushbottom

\section{Introduction}\label{sec:in}
The study of \ac{SGWB} has been ongoing for many years, and various origins have been proposed, including cosmological phase transition \cite{Hindmarsh:2017gnf, Guo:2020grp}, cosmic strings, domain walls \cite{Hiramatsu:2010yz,Blanco-Pillado:2011egf, Kawasaki:2011vv}, \ac{SMBHB} \cite{Rajagopal:1994zj,Kocsis:2010xa}, and \acp{SIGW} \cite{Ananda_2007, Kohri:2018awv}. The latest observations by NANOGrav \cite{NANOGrav:2023gor,NANOGrav:2023hde,NANOGrav:2023hvm}, EPTA \cite{EPTA:2023_1,EPTA:2023_2,EPTA:2023_3,EPTA:2023_4,EPTA:2023_5,EPTA:2023_6}, PPTA \cite{PPTA:2023_1,PPTA:2023_2,PPTA:2023_3}, and CPTA \cite{CPTA:2023} have uncovered evidence of the existence of SGWB and provided greater insight into the origins of the \ac{SGWB}. Extensive research into the origin of SGWB has been conducted through PTA data \cite{Ellis:2023tsl,Kitajima:2023cek,Bai:2023cqj,Fujikura:2023lkn,Bringmann:2023opz,Depta:2023qst,Balaji:2023ehk,Cai:2023dls,wang2023exploring,Vagnozzi:2023lwo,Inomata:2023zup}. With the Bayes factor, it was found that \acp{SIGW} is a possible candidate for the \acp{SGWB} observed by PTA \cite{NANOGrav:2023hvm}. This paper focuses on the \acp{SGWB} dominated by \acp{SIGW}.

During the inflation, the primordial cosmological perturbations were generated by the quantum fluctuations. Based on helicity decomposition of metric perturbations, primordial perturbations can be decomposed as scalar, vector, and tensor perturbations. Since the primordial vector perturbation decays very fast \cite{Saga:2017hft}, it has little influence on the evolution of the Universe. Observations of the \ac{CMB} give very tight limits on scalar and tensor perturbations at large scales, while there are almost no constraints on these perturbations at small scales \cite{Planck:2018jri}. Thus, a sufficiently large energy density spectrum of \acp{SIGW} can be obtained from the power spectra of larger primordial perturbations on the small scales. In addition, unlike CMB, which can only provide the physics imprinted on its last-scattering surface, \acp{SIGW} can carry information with higher redshift. More precisely, the \ac{PTA}, as well as the future observations by \ac{LISA} and Taiji of the \acp{SGWB}, emerge as probes for primordial perturbations generated during inflation. Consequently, we can constrain the parameters of the inflation models in terms of the observation of \acp{SIGW}.

The \acp{SIGW} have been studied for many years as a prediction of cosmological perturbation theory. There are lots of research related to second order \acp{SIGW}, including primordial non-Gaussianity \cite{Adshead:2021hnm, Yuan:2023ofl, Li:2023xtl}, the relationship with \acp{PBH} \cite{Yuan:2021qgz, DeLuca:2023tun}, gauge issue \cite{Hwang:2017oxa, Yuan:2019fwv, Inomata:2019yww, Domenech:2021and}, damping effect \cite{Saga:2014jca, Zhang:2022dgx} and so on. For the third order \acp{SIGW}, Ref.~\cite{Yuan:2019udt} partly considered the third order effects of \acp{SIGW} which are induced by first order
scalar perturbation directly. In Refs.~\cite{Zhou:2021vcw,Chang:2022nzu}, we considered all kinds of source terms and investigated the third order \acp{SIGW} systematically. In Refs.~\cite{Wang:2023sij}, the impact of third order \acp{SIGW} on \ac{CMB}, \ac{BBN}, and \ac{PTA} observations were investigated. However, in previous research on third order \acp{SIGW}, the focus was on monochromatic primordial power spectra characterized by infinitesimal width and infinitely high peak. Under these conditions, third order \acp{SIGW} have a significant impact on the energy density spectrum in the high-frequency region. Notably, when $A_{\zeta}\sim 10^{-2}$, third order \acp{SIGW} exhibit energy density levels comparable to second order \acp{SIGW}. In this paper, we revisit the topic of third order \acp{SIGW}. We explore a more realistic log-normal primordial power spectrum and systematically analyze the contributions of various source terms to the energy density spectrum of third order \acp{SIGW}. We find that the substantial contribution of the third order \acp{SIGW} in the high-frequency region is not solely due to the special nature of monochromatic power spectra, this contribution also persists for more general log-normal primordial power spectra. 

This paper is organized as follows. In Sec.~\ref{sec:2.0}, we review the second and third order \acp{SIGW}. In Sec.~\ref{sec:3.0} we investigate the energy density spectra of third order \acp{SIGW} for the log-normal primordial power spectra. In Sec.~\ref{sec:4.0}, we perform Bayesian analysis by combining \ac{PTA} observational data and investigate the impact of third order \acp{SIGW} on the \ac{SNR} of \ac{LISA}. Finally, we summarize our results and give some discussions in Sec.~\ref{sec:5.0}.

\section{Second and third order \acp{SIGW}}\label{sec:2.0}
In this section, we briefly review the main results of second and third order SIGWs during the \ac{RD} era. The perturbed metric in the flat \ac{FLRW} spacetime with Newtonian gauge is given by
\begin{equation}\label{eq:ds}
\begin{aligned}
	\mathrm{d} s^2 & =a^2(\eta)\left[-\left(1+2 \phi^{(1)}+\phi^{(2)}\right) \mathrm{d} \eta^2+V_i^{(2)} \mathrm{d} \eta \mathrm{d} x^i\right. \\
	& \left.+\left(\left(1-2 \psi^{(1)}-\psi^{(2)}\right) \delta_{i j}+\frac{1}{2} h_{i j}^{(2)}+\frac{1}{6} h_{i j}^{(3)}\right) \mathrm{d} x^i \mathrm{~d} x^j\right] \ ,
\end{aligned}
\end{equation}
where $\eta$ is the conformal time. $\phi^{(n)}$ and $\psi^{(n)}$$\left( n=1,2 \right)$ are first order and second order scalar perturbations. $h^{(n)}_{ij}$$\left( n=2,3 \right)$ are second order and third order tensor perturbations. $V_i^{(2)} $ is second order vector perturbation. Here, we have neglected the first order vector and tensor perturbations. The total contributions of \acp{SIGW} up to third order can be written as
\begin{equation}\label{eq:hp}
\begin{aligned}
	h_{ij}(\mathbf{x},\eta)=\frac{1}{2}h^{(2)}_{ij}(\mathbf{x},\eta)+\frac{1}{6}h^{(2)}_{ij}(\mathbf{x},\eta) \ ,
\end{aligned}
\end{equation}
where $h^{(2)}_{ij}(\mathbf{x},\eta)$ and $h^{(3)}_{ij}(\mathbf{x},\eta)$ are known as second and third order \acp{SIGW} respectively. The Fourier components of $h_{ij}^{(n)}(\mathbf{x},\eta)$$(n=2,3)$ in terms of the polarization tensors $\varepsilon_{i j}^{\lambda}(\mathbf{k})$ $(\lambda=+,\times)$ are defined as
\begin{equation}\label{eq:h}
\begin{aligned}
h_{i j}^{(n)}(\mathbf{x}, \eta)=\int \frac{\mathrm{d}^3 k}{(2 \pi)^{3 / 2}} e^{i \mathbf{k} \cdot \mathbf{x}}\left(h_{\mathbf{k}}^{+,(n)}(\eta) \varepsilon_{i j}^{+}(\mathbf{k})+h_{\mathbf{k}}^{\times,(n)}(\eta) \varepsilon_{i j}^{\times}(\mathbf{k})\right) \ , \ (n=,2,3) \ .
\end{aligned}
\end{equation}
 The explicit expressions of the polarization tensors $\varepsilon_{i j}^{\lambda}(\mathbf{k})$ $(\lambda=+,\times)$ are given in Appendix.~\ref{sec:A}. In Sec.~\ref{sec:2.1}, we will present three second order perturbations that are induced by first order scalar perturbation. We will then discuss third order gravitational waves that are induced by all lower order perturbations. The power spectra of second and third order \acp{SIGW} and corresponding loop diagrams will be investigated in Sec.~\ref{sec:2.2}.

 \subsection{Equations of motion}\label{sec:2.1}
 To calculate the second and third order \acp{SIGW} in Eq.~(\ref{eq:h}), we need to solve the cosmological perturbation equation order by order. The equations of motion of first order scalar perturbations can be written as \cite{Inomata:2020cck}
\begin{equation}\label{eq:phi1}
	\begin{aligned}
		3  \psi^{(1)''}(\mathbf{x},\eta)-\Delta \psi^{(1)}(\mathbf{x},\eta)+3 \mathcal{H}\left( \phi^{(1)'}(\mathbf{x},\eta)+3  \psi^{(1)'}(\mathbf{x},\eta)\right) &=0 \ , \\
		\psi^{(1)}(\mathbf{x},\eta)-\phi^{(1)}(\mathbf{x},\eta) &=0 \ ,
	\end{aligned}
\end{equation}
where the prime denotes the derivative with respect to the conformal time $\eta$. $\mathcal{H}=a'/a$ is the conformal Hubble parameter. By solving Eq.~(\ref{eq:phi1}) during the \ac{RD} era ($\mathcal{H}=1/\eta$, $\omega=1/3$, and $c_s^2=1/3$), we obtain the first order scalar perturbations in momentum space: $\phi(\mathbf{k},\eta)=\psi(\mathbf{k},\eta)=\frac{2}{3}\zeta_{\mathbf{k}} T_{\phi}(|\mathbf{k}| \eta)$, 
where $\zeta_{\mathbf{k}}$ is the primordial curvature perturbation. The transfer function $ T_\phi(|\mathbf{k}| \eta)$ of first order scalar perturbation is given by $T_{\phi}(|\mathbf{k}|\eta)=\frac{9}{(|\mathbf{k}|\eta)^{2}}\left(\frac{\sqrt{3}}{|\mathbf{k}|\eta} \sin \left(\frac{|\mathbf{k}| \eta}{\sqrt{3}}\right)-\cos \left(\frac{|\mathbf{k}| \eta}{\sqrt{3}}\right)\right)$ \cite{Inomata:2020cck}.

The equations of motion of three kinds of second order cosmological perturbations are given by \cite{Inomata:2020cck,Chang:2022dhh,Kohri:2018awv}
\begin{eqnarray}
		&&\psi^{(2)}(\mathbf{x},\eta)-\phi^{(2)}(\mathbf{x},\eta)=-2 \Delta^{-1}\left(\partial^{r} \Delta^{-1} \partial^{s}-\frac{1}{2} \mathcal{T}^{rs}\right) \mathcal{S}^{(2)}_{rs}(\mathbf{x},\eta) \ , \label{eq:2eS1}\\
		&&\psi^{(2)''}(\mathbf{x},\eta)+3\mathcal{H}  \psi^{(2)'}(\mathbf{x},\eta)-\frac{5}{6} \Delta \psi^{(2)}(\mathbf{x},\eta)+\mathcal{H}  \phi^{(2)'}(\mathbf{x},\eta)+\frac{1}{2} \Delta \phi^{(2)}(\mathbf{x},\eta)\nonumber\\
		&&~~~~~~~=-\frac{1}{2} \mathcal{T}^{rs} \mathcal{S}^{(2)}_{rs}(\mathbf{x},\eta) \ , \label{eq:2eS2}\\
        &&V_{l}^{(2)'}(\mathbf{x},\eta)+2 \mathcal{H} V_{l}^{(2)}(\mathbf{x},\eta) =4 \Delta^{-1} \mathcal{T}_{l}^{r} \partial^{s} \mathcal{S}^{(2)}_{rs}(\mathbf{x},\eta) \ , \label{eq:2eV}\\
        &&h_{lm}^{(2)''}(\mathbf{x},\eta)+2 \mathcal{H}  h_{lm}^{(2)'}(\mathbf{x},\eta)-\Delta h_{lm}^{(2)}(\mathbf{x},\eta)=-4 \Lambda_{lm}^{rs} \mathcal{S}^{(2)}_{rs}(\mathbf{x},\eta) \ , \label{eq:2eT}
\end{eqnarray}
where $\Lambda_{rs}^{l m}=\mathcal{T}_{r}^{l} \mathcal{T}_{s}^{m}-\frac{1}{2} \mathcal{T}_{rs} \mathcal{T}^{l m}$ is the transverse and traceless operator, and $\mathcal{T}_{r}^{l}$ is defined as $\mathcal{T}_{r}^{l}=\delta_{r}^{l}-\partial^{l} \Delta^{-1} \partial_{r}$. The source term in Eq.~(\ref{eq:2eS1})--Eq.~(\ref{eq:2eT}) is given by \cite{Chang:2020tji}
\begin{eqnarray}
		\mathcal{S}^{(2)}_{rs}(\mathbf{x},\eta)&&= \partial_{r} \phi^{(1)} \partial_{s} \phi^{(1)}+4 \phi^{(1)} \partial_{r} \partial_{s} \phi^{(1)}
		-\frac{1}{ \mathcal{H}}\left(\partial_{r} \phi^{(1)'} \partial_{s} \phi^{(1)}+\partial_{r} \phi^{(1)} \partial_{s} \phi^{(1)'}\right)\nonumber\\
  &&-\frac{1}{ \mathcal{H}^{2}} \partial_{r}  \phi^{(1)'} \partial_{s}  \phi^{(1)'}-\delta_{ij}\left( \frac{11}{3} \partial_{k} \phi^{(1)} \partial^{k} \phi^{(1)}+24 \mathcal{H} \phi^{(1)}  \phi^{(1)'}+\frac{16}{3} \phi^{(1)} \Delta \phi^{(1)} \right.\nonumber\\
  &&\left.-\frac{2}{3 \mathcal{H}} \partial_{k}  \phi^{(1)'} \partial^{k} \phi^{(1)}+2\left( \phi^{(1)'}\right)^{2}+4 \phi^{(1)} \phi^{(1)''} -\frac{1}{3\mathcal{H}^{2}} \partial_{k} \phi^{(1)'} \partial^{k} \phi^{(1)'} \right) \ .
\end{eqnarray}
Eq.~(\ref{eq:2eS1})--Eq.~(\ref{eq:2eS2}) and Eq.~(\ref{eq:2eV}) represent the equations of motion of second order scalar induced scalar and vector perturbations respectively. The equation of motion of second order \acp{SIGW} is given in Eq.~(\ref{eq:2eT}). By solving Eq.~(\ref{eq:2eS1})--Eq.~(\ref{eq:2eT}), we obtain the second order perturbations in momentum space
\begin{eqnarray}
		\psi^{(2)}(\mathbf{k},\eta)&=&\int \frac{\mathrm{d}^{3} q}{(2 \pi)^{3/2}}  I_{\psi}^{(2)}(|\mathbf{k}-\mathbf{p}|,|\mathbf{p}|,\eta)\zeta_{\mathbf{k}-\mathbf{p}} \zeta_{\mathbf{p}} \ , \\
  \phi^{(2)}(\mathbf{k},\eta)&=&\int \frac{\mathrm{d}^{3} q}{(2 \pi)^{3/2}}  I_{\phi}^{(2)}(|\mathbf{k}-\mathbf{p}|,|\mathbf{p}|,\eta)\zeta_{\mathbf{k}-\mathbf{p}} \zeta_{\mathbf{p}} \ , \\
V^{\lambda,(2)}(\mathbf{k},\eta)&=&\int \frac{d^3 p}{(2 \pi)^{3 / 2}}  \frac{ik^s e^{\lambda, r}(\mathbf{k})}{k^2} p_r p_s I_V^{(2)}(|\mathbf{k}-\mathbf{p}|,|\mathbf{p}|,\eta)\zeta_{\mathbf{k}-\mathbf{p}} \zeta_{\mathbf{p}} \ , \\
  	h^{\lambda,(2)}(\mathbf{k},\eta)&=&\int \frac{d^3 p}{(2 \pi)^{3 / 2}}  \varepsilon^{\lambda, l m}(\mathbf{k})p_l p_m  I_{h}^{(2)}(|\mathbf{k}-\mathbf{p}|,|\mathbf{p}|,\eta)\zeta_{\mathbf{k}-\mathbf{p}}  \zeta_{\mathbf{p}}  \ , \label{eq:2h0} 
\end{eqnarray}
where $I^{(2)}_i$$(i=\phi,\psi,V,h)$ are second order kernel functions. The analytical expressions of second order kernel functions can be found in Refs.~\cite{Zhou:2021vcw}.

After calculating all first order and second order perturbations, we can systematically study the third order \acp{SIGW}. The equation of motion of third order \acp{SIGW} is as follows
\begin{equation}\label{eq:eh3}
	h_{i j}^{(3)''}(\mathbf{x},\eta)+2 \mathcal{H}  h_{i j}^{(3)'}(\mathbf{x},\eta)-\Delta h_{i j}^{(3)}(\mathbf{x},\eta)=-12 \Lambda_{i j}^{l m} \mathcal{S}^{(3)}_{l m}(\mathbf{x},\eta) \ ,
\end{equation}
where the third order source term $\mathcal{S}^{(3)}_{l m}$ can be divided into four parts
\begin{equation}\label{eq:S}
	\mathcal{S}_{lm}^{(3)}(\mathbf{x},\eta)=\mathcal{S}_{lm,\phi\phi\phi}^{(3)}(\mathbf{x},\eta)+\mathcal{S}_{lm,\phi h_{\phi\phi}}^{(3)}(\mathbf{x},\eta)+\mathcal{S}_{lm,\phi V_{\phi\phi}}^{(3)}(\mathbf{x},\eta)+\mathcal{S}_{lm,\phi\psi_{\phi\phi}}^{(3)}(\mathbf{x},\eta) \ .
\end{equation}
The subscripts in Eq.~(\ref{eq:S}) represent the source of \acp{SIGW}. For example, the symbol $\mathcal{S}^{(3)}_{\phi V_{\phi\phi}}$ represents the source term of first order scalar perturbation $\phi$ and second order scalar induced vector perturbation $V_{\phi\phi}$. The explicit expressions of four kinds of source terms in Eq.~(\ref{eq:S}) are given by \cite{Zhou:2021vcw}
\begin{eqnarray}\label{eq:3s}
		S_{lm,\phi\phi\phi}^{(3)}(\mathbf{x},\eta)&=&12\phi^{(1)}\partial_l\phi^{(1)}\partial_m\phi^{(1)}-\frac{4}{\mathcal{H}}\phi^{(1)'}\partial_l\phi^{(1)}\partial_m\phi^{(1)}+\frac{2}{3\mathcal{H}^2}\Delta\phi^{(1)}\partial_l\phi^{(1)}\partial_m\phi^{(1)}\nonumber\\
		&+&\frac{2}{3\mathcal{H}^4}\Delta\phi^{(1)}\partial_l\phi^{(1)'}\partial_m\phi^{(1)'}-\frac{3}{\mathcal{H}^2}\phi^{(1)'}\partial_l\phi^{(1)'}\partial_m\phi^{(1)}-\frac{3}{\mathcal{H}^2}\phi^{(1)'}\partial_m\phi^{(1)'}\partial_l\phi^{(1)}\nonumber\\
		&+&\frac{2}{3\mathcal{H}^3}\Delta\phi^{(1)}\partial_l\phi^{(1)'}\partial_m\phi^{(1)}+\frac{2}{3\mathcal{H}^3}\Delta\phi^{(1)}\partial_m\phi^{(1)'}\partial_l\phi^{(1)}\nonumber\\
		&-&\frac{2}{\mathcal{H}^3}\phi^{(1)'}\partial_l\phi^{(1)'}\partial_m\phi^{(1)'}-\frac{4}{\mathcal{H}^2}\phi^{(1)}\partial_l\phi^{(1)'}\partial_m\phi^{(1)'} \ ,\\
		S_{lm,\phi h_{\phi\phi}}^{(3)}(\mathbf{x},\eta)&=&-\frac{1}{2}\phi^{(1)}\left( h_{lm}^{(2)''}+2 \mathcal{H}  h_{lm}^{(2)'}-\Delta h_{lm}^{(2)}\right)-\phi^{(1)}\Delta h_{lm}^{(2)}-\phi^{(1)'}\mathcal{H}h_{lm}^{(2)}-\frac{1}{3}\Delta \phi^{(1)}h_{lm}^{(2)}\nonumber\\
		&-&\partial^b \phi^{(1)}\partial_b h_{lm}^{(2)} \ , \\
		S_{lm,\phi V_{\phi\phi}}^{(3)}(\mathbf{x},\eta)&=&\phi^{(1)}\partial_l\left(V_m^{(2)'}+2 \mathcal{H}V_m^{(2)} \right)+\phi^{(1)}\partial_m\left(V_l^{(2)'}+2 \mathcal{H}V_l^{(2)} \right)+\phi^{(1)'}\left(\partial_lV_m^{(2)}+\partial_mV_l^{(2)}\right)\nonumber\\
		&-&\frac{\phi^{(1)}}{8\mathcal{H}}\left(\partial_m\Delta V_l^{(2)}+\partial_l\Delta V_m^{(2)}\right)-\frac{\phi^{(1)'}}{8\mathcal{H}^2}\left(\partial_m\Delta V_l^{(2)}+\partial_l\Delta V_m^{(2)}\right) \ , \\
		S_{lm,\phi \psi_{\phi\phi}}^{(3)}(\mathbf{x},\eta)&=&\frac{1}{\mathcal{H}}\left(\phi^{(1)}\partial_l\partial_m\psi^{(2)'}\right)+\frac{1}{\mathcal{H}}\left(\phi^{(1)'}\partial_l\partial_m\phi^{(2)}\right)+\frac{1}{\mathcal{H}^2}\left(\phi^{(1)'}\partial_l\partial_m\psi^{(2)'}\right)\nonumber\\
		&+&3\left(\phi^{(1)}\partial_l\partial_m\phi^{(2)}\right) \ .
\end{eqnarray}
By solving Eq.~(\ref{eq:eh3}), we obtain the third order \acp{SIGW} in momentum space \cite{Chang:2022nzu}
\begin{eqnarray}\label{eq:3h}
h^{\lambda,(3)}(\mathbf{k},\eta)=
h_{ \phi \phi \phi}^{\lambda,(3)}(\mathbf{k},\eta)+
h_{ \phi h_{\phi \phi}}^{\lambda,(3)}(\mathbf{k},\eta)+
h_{ \phi V_{\phi \phi}}^{\lambda,(3)}(\mathbf{k},\eta)+
h_{ \phi \psi_{\phi \phi}}^{\lambda,(3)}(\mathbf{k},\eta) \  ,
\end{eqnarray}
where
\begin{eqnarray}
h_{\phi \phi \phi}^{\lambda,(3)}(\mathbf{k},\eta)&=&\int \frac{\mathrm{d}^3 p}{(2 \pi)^{3 / 2}} \int \frac{\mathrm{d}^3 q}{(2 \pi)^{3 / 2}} \varepsilon^{\lambda, l m}(\mathbf{k}) q_m \left(p_l-q_l\right) \nonumber\\
&\times&  I_{\phi \phi \phi}^{(3)}(u, \bar{u}, \bar{v}, x) \zeta_{\mathbf{k}-\mathbf{p}} \zeta_{\mathbf{p}-\mathbf{q}} \zeta_{\mathbf{q}}\ , \label{eq:3h1}\\
h_{\phi h_{\phi \phi}}^{\lambda,(3)}(\mathbf{k},\eta)&=&\int \frac{\mathrm{d}^3 p}{(2 \pi)^{3 / 2}} \int \frac{\mathrm{d}^3 q}{(2 \pi)^{3 / 2}} \varepsilon^{\lambda, l m}(\mathbf{k}) \Lambda_{l m}^{r s}(\mathbf{p}) q_r q_s \nonumber\\
&\times& I_{\phi h_{\phi \phi}}^{(3)}(u, \bar{u}, \bar{v}, x) \zeta_{\mathbf{k}-\mathbf{p}} \zeta_{\mathbf{p}-\mathbf{q}} \zeta_{\mathbf{q}}\ , \label{eq:3h2}\\
h_{ \phi V_{\phi \phi}}^{\lambda,(3)}(\mathbf{k},\eta)&=&\int \frac{\mathrm{d}^3 p}{(2 \pi)^{3 / 2}} \int \frac{\mathrm{d}^3 q}{(2 \pi)^{3 / 2}} \varepsilon^{\lambda, l m}(\mathbf{k}) \left(\mathcal{T}_{m}^r(\mathbf{p}) p_{l}+\mathcal{T}_{l}^r(\mathbf{p}) p_{m}\right)  \frac{p^s}{ p^2} q_r q_s \nonumber\\
&\times& I_{\phi V_{\phi \phi}}^{(3)}(u, \bar{u}, \bar{v}, x) \zeta_{\mathbf{k}-\mathbf{p}} \zeta_{\mathbf{p}-\mathbf{q}} \zeta_{\mathbf{q}}\ , \label{eq:3h3}\\
h_{ \phi \psi_{\phi \phi}}^{\lambda,(3)}(\mathbf{k},\eta)&=&\int \frac{\mathrm{d}^3 p}{(2 \pi)^{3 / 2}} \int \frac{\mathrm{d}^3 q}{(2 \pi)^{3 / 2}} \varepsilon^{\lambda, l m}(\mathbf{k}) p_l p_m \nonumber\\
&\times&  I_{\phi \psi_{\phi \phi}}^{(3)}(u, \bar{u}, \bar{v}, x) \zeta_{\mathbf{k}-\mathbf{p}} \zeta_{\mathbf{p}-\mathbf{q}} \zeta_{\mathbf{q}} \  . \label{eq:3h4}
\end{eqnarray}
The explicit expressions of third order kernel functions in Eq.~(\ref{eq:3h1})--Eq.~(\ref{eq:3h4}) can be found in Refs.~\cite{Zhou:2021vcw,Chang:2022nzu}.

\subsection{Power spectra of \acp{SIGW}}\label{sec:2.2}
\begin{figure*}[htbp]
    \captionsetup{
      justification=raggedright,
      singlelinecheck=true
    }
    \centering

	\subfloat[$\langle \zeta_{\mathbf{k}-\mathbf{p}}\zeta_{\mathbf{k}'-\mathbf{p}'} \rangle \langle \zeta_{\mathbf{p}}\zeta_{\mathbf{p}'} \rangle$]{\includegraphics[width=.3\columnwidth]{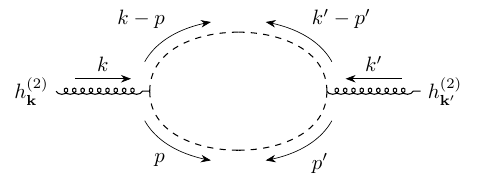}}
        \subfloat[$\langle \zeta_{\mathbf{k}-\mathbf{p}}\zeta_{\mathbf{p}'} \rangle \langle \zeta_{\mathbf{p}}\zeta_{\mathbf{k}'-\mathbf{p}'} \rangle$]{\includegraphics[width=.3\columnwidth]{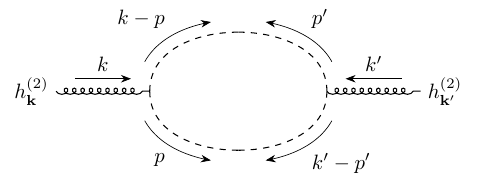}}
        \subfloat[$\langle \zeta_{\mathbf{k}-\mathbf{p}}\zeta_{\mathbf{p}} \rangle \langle \zeta_{\mathbf{k}'-\mathbf{p}'} \zeta_{\mathbf{p}'} \rangle$]{\includegraphics[width=.35\columnwidth]{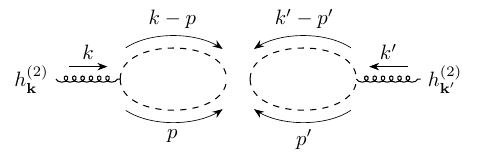}} 
        
\caption{\label{fig:FeynDiag22} The diagrams correspond to the Wick’s expansions of four-point correlation function. The dashed and spring-like lines in the figure represent scalar and tensor perturbations, respectively. Diagram (a) and diagram (b) correspond to the nontrivial contributions of the power spectrum of second order \acp{SIGW} in Eq.~(\ref{eq:p20}).}
\end{figure*}

The power spectra of $n$-th order \acp{GW} $\mathcal{P}_{h}^{(n)}( \mathbf{k},\eta)$ are defined as
\begin{equation}\label{eq:Ph}
  \left\langle h^{\lambda,(n)}( \mathbf{k},\eta) h^{\lambda^{\prime},(n)}\left(\mathbf{k}^{\prime},\eta\right)\right\rangle=\delta^{\lambda \lambda^{\prime}} \delta\left(\mathbf{k}+\mathbf{k}^{\prime}\right) \frac{2 \pi^2}{k^3} \mathcal{P}_{h}^{(n)}(k,\eta) \ .  
\end{equation}
As shown in Eq.~(\ref{eq:Ph}), the power spectrum of \acp{GW} can be calculated in terms of the corresponding two-point function. By substituting Eq.~(\ref{eq:2h0}) into Eq.~(\ref{eq:Ph}) and simplifying the momentum integral, we obtain the explicit expression of the power spectra of second order \acp{SIGW}
\begin{equation}
\begin{aligned}\label{eq:p20}
  \mathcal{P}_{h}^{(2)}(k,\eta)&=4 \int_0^{\infty} \mathrm{d} v \int_{|1-v|}^{1+v} \mathrm{~d} u\left(\frac{4 v^2-\left(1+v^2-u^2\right)^2}{4 v u}\right)^2 \left(I^{(2)}_h(v, u, x)\right)^2 \mathcal{P}_\zeta(k v) \mathcal{P}_\zeta(k u) \ , 
   \end{aligned}
\end{equation}
where we have set $|\mathbf{k}-\mathbf{p}|=u|\mathbf{k}|$, $|\mathbf{p}|=v|\mathbf{k}|$, and $x=|\mathbf{k}|\eta$. $P_\zeta(k)$ is the power spectrum of primordial curvature perturbation which is defined as $\left\langle\zeta_{\mathbf{k}} \zeta_{\mathbf{k}^{\prime}}\right\rangle=\frac{2 \pi^2}{k^3} \delta\left(\mathbf{k}+\mathbf{k}^{\prime}\right) \mathcal{P}_\zeta(k)$. As shown in Fig.~\ref{fig:FeynDiag22}, the two-point function of second order \acp{SIGW} corresponds to two equivalent one-loop diagrams in cosmological perturbation theory. 

Similarly, the explicit expression of the power spectra of third order \acp{SIGW} can be obtained by using Eq.~(\ref{eq:3h}) and Eq.~(\ref{eq:Ph})
\begin{equation}\label{eq:P30}
\begin{aligned}
   \mathcal{P}_{h}^{(3)}(k,\eta)&=\sum_{i,j}\mathcal{P}_{ij}^{(3)} \sim \sum_{i,j}~\langle  h^{\lambda,(3)}_i( \mathbf{k},\eta) h_j^{\lambda^{\prime},(3)}\left(\mathbf{k}^{\prime},\eta\right) \rangle \ , \ (i,j=\phi \phi \phi,~\phi h_{\phi \phi},~\phi V_{\phi \phi},~\phi \psi_{\phi \phi}) \ ,
   \end{aligned}
\end{equation}
where the subscripts $i,j$ in Eq.~(\ref{eq:P30}) represent the four kinds of source terms of third order \acp{SIGW} in Eq.~(\ref{eq:3h1})--Eq.~(\ref{eq:3h4}). More precisely, we will encounter the following sixteen terms in Eq.~(\ref{eq:P30}) when calculating the power spectra of third order \acp{SIGW} \cite{Zhou:2021vcw}
\begin{equation}\label{eq:P3}
\begin{aligned}
\mathcal{P}_{ij}^{(3)} &=\frac{1}{2 \pi} \int_0^{\infty} \mathrm{d} v \int_0^{\infty} \mathrm{d} \bar{v} \int_{|1-\bar{v} v|}^{1+\bar{v} v} \mathrm{~d} w \int_{|1-v|}^{1+v} \mathrm{~d} u \int_{\bar{u}_{-}}^{\bar{u}_{+}} \mathrm{d} \bar{u}\left\{\frac{w}{u^2 \bar{u}^2 v \bar{v}^2 \sqrt{Y\left(1-X^2\right)}} \right. \\
& \left.\times I_i^{(3)}(u, v, \bar{u}, \bar{v}, x) \sum^{6}_{a=1} ~\mathbb{P}^{a}_{i j}(\mathbf{k},\mathbf{p}, \mathbf{q})I_{j,a}^{(3)}(u, v, \bar{u}, \bar{v},w, x) P_{\zeta}(k u) P_{\zeta}(k \bar{u} v) P_{\zeta}(k \bar{v} v)\right\}\ , \\
& (i,j=\phi \phi \phi,~\phi h_{\phi \phi},~\phi V_{\phi \phi},~\phi \psi_{\phi \phi}) \ ,
\end{aligned}
\end{equation}
where
\begin{equation}
\begin{aligned}
\bar{u}_{ \pm}&=\left(1+\bar{v}^2-\frac{\left(1+v^2-u^2\right)\left(1+(\bar{v} v)^2-w^2\right)}{2 v^2}\right. \\
&\left. \pm 2 \bar{v} \sqrt{\left(1-\left(\frac{1+v^2-u^2}{2 v}\right)^2\right)\left(1-\left(\frac{1+(\bar{v} v)^2-w^2}{2 \bar{v} v}\right)^2\right)}\right)^{\frac{1}{2}} \ , \\
X&=\left(-1+u^2+v^2-2 \bar{u}^2 v^2+v^2 \bar{v}^2+u^2 v^2 \bar{v}^2-v^4 \bar{v}^2+w^2-u^2 w^2+v^2 w^2\right) \\
&\times\left(1-2 u^2+u^4-2 v^2-2 u^2 v^2+v^4\right)\left(1-2 v^2 \bar{v}^2+v^4 \bar{v}^4-2 w^2-2 v^2 \bar{v}^2 w^2+w^4\right)^{-\frac{1}{2}} \ , \\
Y&=\left(1-2 u^2+u^4-2 v^2-2 u^2 v^2+v^4\right)\left(1-2 v^2 \bar{v}^2+v^4 \bar{v}^4-2 w^2-2 v^2 \bar{v}^2 w^2+w^4\right) \ .
\end{aligned}
\end{equation}
In Eq.~(\ref{eq:P3}), we have set $|\mathbf{p}-\mathbf{q}|=\bar{u}|\mathbf{p}|$, $|\mathbf{k}-\mathbf{q}|=w|\mathbf{k}|$, and $|\mathbf{q}|=\bar{v}|\mathbf{p}|$. The corresponding two-loop diagrams in cosmological perturbation theory are given in Appendix.~\ref{sec:C}.

The summation of index $a$ in Eq.~(\ref{eq:P3}) origins from the Wick’s expansions of the six-point function of $\zeta_{\mathbf{k}}$ in Appendix.~\ref{sec:B}. More precisely, the sum of the six terms ($a=1,\cdot\cdot\cdot,6$) in Eq.~(\ref{eq:P3}) corresponds to the six terms in the Wick's expansions in Eq.~(\ref{eq:Wick}). After integrating the three-dimensional delta functions in the Wick's theorem expansion, we can replace the momenta $\mathbf{k}'$, $\mathbf{p}'$, and $\mathbf{q}'$ with linear combinations of momenta $\mathbf{k}$, $\mathbf{p}$, and $\mathbf{q}$. In Eq.~(\ref{eq:P3}), the two multiplied kernel functions can be represented as $I_i^{(3)}(|\mathbf{k}-\mathbf{p}|,|\mathbf{p}-\mathbf{q}|,|\mathbf{q}|,\eta)$ and $I_j^{(3)}(|\mathbf{k}'-\mathbf{p}'|,|\mathbf{p}'-\mathbf{q}'|,|\mathbf{q}'|,\eta)$.  It can be observed that since the parameters of the kernel function $I_i^{(3)}(|\mathbf{k}-\mathbf{p}|,|\mathbf{p}-\mathbf{q}|,|\mathbf{q}|,\eta)$ only contain $\mathbf{k}$, $\mathbf{p}$, and $\mathbf{q}$, the variable replacement in the Wick expansion will only affect the kernel function $I_j^{(3)}$. Therefore, the kernel function $I_j^{(3)}$ needs to be transformed into $I_{j,a}^{(3)}$ and summed over the upper index $a$, which corresponds to summing over the six terms in Wick's expansion. The detailed discussions can be found in Refs.~\cite{Zhou:2021vcw,Chang:2022dhh}.

The momentum polynomials $\mathbb{P}^{a}_{i j}(\mathbf{k},\mathbf{p}, \mathbf{q})$ in Eq.~(\ref{eq:P3}) can be calculated in terms of the formal expressions of third order \acp{SIGW} in Eq.~(\ref{eq:3h1})--Eq.~(\ref{eq:3h4}). The explicit expressions of $\mathbb{P}^{a}_{i j}(\mathbf{k},\mathbf{p}, \mathbf{q})$ for $i,j=\phi \phi \phi,~\phi h_{\phi \phi},~\phi V_{\phi \phi}$, and $\phi \psi_{\phi \phi}$ are given in Table.~\ref{ta:1}. Similarly, the superscript $a=1,\cdots,6$ in $\mathbb{P}^{a}_{i j}$ represents the six terms in Wick's expansion. The momentum polynomials $\mathbb{P}^{a}_{i j}$ in Table.~\ref{ta:1} can be calculated directly in a given coordinate system using the expression for the polarization tensor in Appendix.~\ref{sec:A}. It should be emphasized that it can also be calculated without selecting a specific coordinate system by using the relationship between the polarization tensor $\varepsilon^{\lambda, l m}(\mathbf{k})$ and the transverse traceless operator $\Lambda_{rs}^{l m}(\mathbf{k})$. Specifically, the polarization tensor satisfies the following expression
\begin{eqnarray}\label{eq:Lam}
    \delta^{\lambda\bar{\lambda}}\varepsilon^{\bar{\lambda}, lm}(\mathbf{k})\varepsilon^{\bar{\lambda}}{ij}(\mathbf{k})=\Lambda_{i j}^{lm}(\mathbf{k})+A_{i j}^{lm}(\mathbf{k}) \ ,
\end{eqnarray}
where $A_{i j}^{lm}(\mathbf{k})$ is defined as
\begin{equation}\label{eq:Aij}
A_{i j}^{lm}(\mathbf{k})\equiv\frac{1}{\sqrt{2}}\left(\bar{e}_i(\mathbf{k})e_j(\mathbf{k})-e_i(\mathbf{k})\bar{e}_j(\mathbf{k}) \right)\frac{1}{\sqrt{2}}\left(\bar{e}_m(\mathbf{k})e_l(\mathbf{k})-e_m(\mathbf{k})\bar{e}_l(\mathbf{k}) \right) \  .
\end{equation}
Since $A_{i j}^{lm}(\mathbf{k})$ is dependent on the polarization vectors $\bar{e}_l(\mathbf{k})$ and $e_l(\mathbf{k})$, it appears that we still need to select a specific coordinate system when calculating the contraction of the polarization tensor with respect to $\lambda$ and $\bar{\lambda}$. However, as shown in Eq.~(\ref{eq:Aij}), $A_{i j}^{lm}(\mathbf{k})$ is an antisymmetric tensor, it satisfies $A_{i j}^{lm}T_{lm}=A_{i j}^{lm}T^{ij}=0$ for arbitrary symmetric tensor $T_{ij}$. Therefore, when $A_{i j}^{lm}(\mathbf{k})$ acts on a symmetric tensor, we obtain
\begin{eqnarray}\label{eq:LA}
    & &\delta^{\lambda\bar{\lambda}}\varepsilon^{\bar{\lambda}, lm}(\mathbf{k})\varepsilon^{\bar{\lambda}}{ij}(\mathbf{k})=\Lambda_{i j}^{lm}(\mathbf{k})=\mathcal{T}_{i}^{l} \mathcal{T}_{j}^{m}-\frac{1}{2} \mathcal{T}_{ij} \mathcal{T}^{l m} \nonumber\\
   & &=\left(\delta_{i}^{l}-\frac{k^{l} k_{i}}{|\mathbf{k}|^{2}}\right)\left(\delta_{j}^{m}-\frac{k^{m} k_{j}}{|\mathbf{k}|^{2}}\right)-\frac{1}{2}\left(\delta_{i j}-\frac{k_{i} k_{j}}{|\mathbf{k}|^{2}}\right)\left(\delta^{l m}-\frac{k^{l} k^{m}}{|\mathbf{k}|^{2}}\right) \ .
\end{eqnarray}
As shown in Eq.~(\ref{eq:LA}), we can calculate the contraction of the polarization tensor with respect to $\lambda$ and $\bar{\lambda}$ without specifying a particular coordinate system. This allows us to simplify the momentum polynomials $\mathbb{P}^{a}_{i j}(\mathbf{k},\mathbf{p}, \mathbf{q})$ without selecting a coordinate system. However, it is important to note that this method is only applicable when $A_{i j}^{lm}(\mathbf{k})$ acts on a symmetric tensor.
\begin{table}[h!]
\centering
\begin{tabular}{|c|c|c|c|c|}
\hline 
\hline
$\mathbb{P}^a_{i j}$ & $i=\phi\phi\phi$   \\
\hline
$j=\phi\phi\phi$ & $\delta^{\lambda \lambda'} \left( \varepsilon^{\lambda, l m}(\mathbf{k}) q_m \left(p_l-q_l\right) \right)\left( \varepsilon^{\lambda', rs}(\mathbf{k}') q'_s \left(p'_r-q'_r\right) \right)$   \\
\hline
$j=\phi h_{\phi \phi}$ & $\delta^{\lambda \lambda'}\left( \varepsilon^{\lambda, l m}(\mathbf{k}) q_m \left(p_l-q_l\right) \right)\left( \varepsilon^{\lambda', bc}(\mathbf{k}') \Lambda_{bc}^{r s}(\mathbf{p}') q'_r q'_s \right)$  \\
\hline
$j=\phi V_{\phi \phi}$ & $\delta^{\lambda \lambda'}\left( \varepsilon^{\lambda, l m}(\mathbf{k}) q_m \left(p_l-q_l\right) \right)\left( \varepsilon^{\lambda', bc}(\mathbf{k}') \left(\mathcal{T}_{c}^r(\mathbf{p}') p'_{b}+\mathcal{T}_{b}^r(\mathbf{p}') p'_{c}\right)  \frac{p'^s}{ p'^2} q'_r q'_s  \right)$  \\
\hline
$j=\phi \psi_{\phi \phi}$ & $\delta^{\lambda \lambda'}\left( \varepsilon^{\lambda, l m}(\mathbf{k}) q_m \left(p_l-q_l\right) \right)\left( \varepsilon^{\lambda', rs}(\mathbf{k}') p'_r p'_s \right)$  \\
\hline
\hline
\hline
$\mathbb{P}^a_{i j}$ & $i=\phi h_{\phi\phi}$   \\
\hline
$j=\phi\phi\phi$ & $\delta^{\lambda \lambda'}\left( \varepsilon^{\lambda, lm}(\mathbf{k}) \Lambda_{lm}^{ef}(\mathbf{p}) q_e q_f \right)\left( \varepsilon^{\lambda', rs}(\mathbf{k}') q'_s \left(p'_r-q'_r\right) \right)$   \\
\hline
$j=\phi h_{\phi \phi}$ & $\delta^{\lambda \lambda'}\left( \varepsilon^{\lambda, lm}(\mathbf{k}) \Lambda_{lm}^{ef}(\mathbf{p}) q_e q_f \right)\left( \varepsilon^{\lambda', bc}(\mathbf{k}') \Lambda_{bc}^{r s}(\mathbf{p}') q'_r q'_s \right)$  \\
\hline
$j=\phi V_{\phi \phi}$ & $\delta^{\lambda \lambda'}\left( \varepsilon^{\lambda, lm}(\mathbf{k}) \Lambda_{lm}^{ef}(\mathbf{p}) q_e q_f \right)\left( \varepsilon^{\lambda', bc}(\mathbf{k}') \left(\mathcal{T}_{c}^r(\mathbf{p}') p'_{b}+\mathcal{T}_{b}^r(\mathbf{p}') p'_{c}\right)  \frac{p'^s}{ p'^2} q'_r q'_s  \right)$  \\
\hline
$j=\phi \psi_{\phi \phi}$ & $\delta^{\lambda \lambda'}\left( \varepsilon^{\lambda, lm}(\mathbf{k}) \Lambda_{lm}^{ef}(\mathbf{p}) q_e q_f \right)\left( \varepsilon^{\lambda', rs}(\mathbf{k}') p'_r p'_s \right)$  \\
\hline
\hline
\hline
$\mathbb{P}^a_{i j}$ & $i=\phi V_{\phi\phi}$   \\
\hline
$j=\phi\phi\phi$ & $\delta^{\lambda \lambda'}\left( \varepsilon^{\lambda, lm}(\mathbf{k}) \left(\mathcal{T}_{m}^e(\mathbf{p}) p_{l}+\mathcal{T}_{l}^e(\mathbf{p}) p_{m}\right)  \frac{p^f}{ p2} q_e q_f  \right)\left( \varepsilon^{\lambda', rs}(\mathbf{k}') q'_s \left(p'_r-q'_r\right) \right)$   \\
\hline
$j=\phi h_{\phi \phi}$ & $\delta^{\lambda \lambda'}\left( \varepsilon^{\lambda, lm}(\mathbf{k}) \left(\mathcal{T}_{m}^e(\mathbf{p}) p_{l}+\mathcal{T}_{l}^e(\mathbf{p}) p_{m}\right)  \frac{p^f}{ p2} q_e q_f  \right)\left( \varepsilon^{\lambda', bc}(\mathbf{k}') \Lambda_{bc}^{r s}(\mathbf{p}') q'_r q'_s \right)$  \\
\hline
$j=\phi V_{\phi \phi}$ & $\delta^{\lambda \lambda'}\left( \varepsilon^{\lambda, lm}(\mathbf{k}) \left(\mathcal{T}_{m}^e(\mathbf{p}) p_{l}+\mathcal{T}_{l}^e(\mathbf{p}) p_{m}\right)  \frac{p^f}{ p2} q_e q_f  \right)\left( \varepsilon^{\lambda', bc}(\mathbf{k}') \left(\mathcal{T}_{c}^r(\mathbf{p}') p'_{b}+\mathcal{T}_{b}^r(\mathbf{p}') p'_{c}\right)  \frac{p'^s}{ p'^2} q'_r q'_s  \right)$  \\
\hline
$j=\phi \psi_{\phi \phi}$ & $\delta^{\lambda \lambda'}\left( \varepsilon^{\lambda, lm}(\mathbf{k}) \left(\mathcal{T}_{m}^e(\mathbf{p}) p_{l}+\mathcal{T}_{l}^e(\mathbf{p}) p_{m}\right)  \frac{p^f}{ p2} q_e q_f  \right)\left( \varepsilon^{\lambda', rs}(\mathbf{k}') p'_r p'_s \right)$  \\
\hline
\hline
\hline
$\mathbb{P}^a_{i j}$ & $i=\phi \psi_{\phi\phi}$   \\
\hline
$j=\phi\phi\phi$ & $\delta^{\lambda \lambda'}\left( \varepsilon^{\lambda, lm}(\mathbf{k}) p_l p_m \right)\left( \varepsilon^{\lambda', rs}(\mathbf{k}') q'_s \left(p'_r-q'_r\right) \right)$   \\
\hline
$j=\phi h_{\phi \phi}$ & $\delta^{\lambda \lambda'}\left( \varepsilon^{\lambda, lm}(\mathbf{k}) p_l p_m \right)\left( \varepsilon^{\lambda', bc}(\mathbf{k}') \Lambda_{bc}^{r s}(\mathbf{p}') q'_r q'_s \right)$  \\
\hline
$j=\phi V_{\phi \phi}$ & $\delta^{\lambda \lambda'}\left( \varepsilon^{\lambda, lm}(\mathbf{k}) p_l p_m \right)\left( \varepsilon^{\lambda', bc}(\mathbf{k}') \left(\mathcal{T}_{c}^r(\mathbf{p}') p'_{b}+\mathcal{T}_{b}^r(\mathbf{p}') p'_{c}\right)  \frac{p'^s}{ p'^2} q'_r q'_s  \right)$  \\
\hline
$j=\phi \psi_{\phi \phi}$ & $\delta^{\lambda \lambda'}\left( \varepsilon^{\lambda, lm}(\mathbf{k}) p_l p_m \right)\left( \varepsilon^{\lambda', rs}(\mathbf{k}') p'_r p'_s \right)$  \\
\hline
\end{tabular}
\caption{The explicit expressions of momentum polynomials $\mathbb{P}^a_{i j}(\mathbf{k},\mathbf{p}, \mathbf{q})$. }
\label{ta:1}
\end{table}

\section{Energy density spectrum with lognormal primordial power spectrum}\label{sec:3.0}
The energy density spectrum of \acp{SIGW} is defined as
\begin{equation}
    \overline{\Omega}_{\mathrm{GW}}(\eta, k) = \overline{\frac{\rho_{\mathrm{GW}}(\eta,k)}{\rho_{\mathrm{tot}}(\eta)}} = \overline{\frac{x^2}{6}\mathcal{P}_h(\eta,k)}  \ ,
\end{equation}
where
\begin{equation}
    \mathcal{P}_h(\eta,k) = \frac{1}{4}\mathcal{P}^{(2)}_h(\eta,k) + \frac{1}{36}\mathcal{P}^{(3)}_h(\eta,k) + O(\mathcal{P}^{(4)}_h) \  ,
\end{equation}
is the total power spectrum of \acp{SIGW}. The overline represents the oscillation average. The total energy density spectrum of \acp{SIGW} up 
to third order can be represented as
\begin{equation}
\begin{aligned}
    \bar{\Omega}^{\mathrm{tot}}_{\mathrm{GW}}(\eta,k)& = \bar{\Omega}^{(2)}_{\mathrm{GW}}(\eta,k) +\bar{\Omega}_{\mathrm{GW}}^{(3)}(\eta,k) \\
    &=\bar{\Omega}^{(2)}_{\mathrm{GW}}(\eta,k) + \sum\limits_{i,j}\bar{\Omega}^{(3)}_{ij}(\eta,k) \ , \ (i,j=\phi \phi \phi,~\phi h_{\phi \phi},~\phi V_{\phi \phi},~\phi \psi_{\phi \phi}) \ ,
    \end{aligned}
\end{equation}
where $\bar{\Omega}_{\mathrm{GW}}^{(2)}(\eta,k)$ and $\bar{\Omega}_{\mathrm{GW}}^{(3)}(\eta,k)$ represent the energy density spectra of second and third order \acp{SIGW}, respectively. Taking into account the thermal history of the universe, we obtain the current total energy density spectrum \cite{Wang:2019kaf}
\begin{equation}
    \bar{\Omega}^{\mathrm{tot}}_{\mathrm{GW,0}}(k) = \Omega_{\mathrm{rad},0}\left(\frac{g_{*,\rho,e}}{g_{*,\rho,0}}\right)\left(\frac{g_{*,s,0}}{g_{*,s,e}}\right)^{4/3}\bar{\Omega}^{\mathrm{tot}}_{\mathrm{GW}}(\eta,k) \ ,
\end{equation}
where $\Omega_{\mathrm{rad},0}$ ($ =4.2\times 10^{-5}h^{-2}$) is the energy density fraction of radiations today, and the dimensionless Hubble constant is $h = 0.6736$ \cite{Planck:2018vyg}. The effect numbers of relativistic species $g_{*,\rho}$ and $g_{*,s}$ can be found in Ref.~\cite{Saikawa:2018rcs}.
\begin{figure}
    \centering
    \includegraphics[scale=0.57]{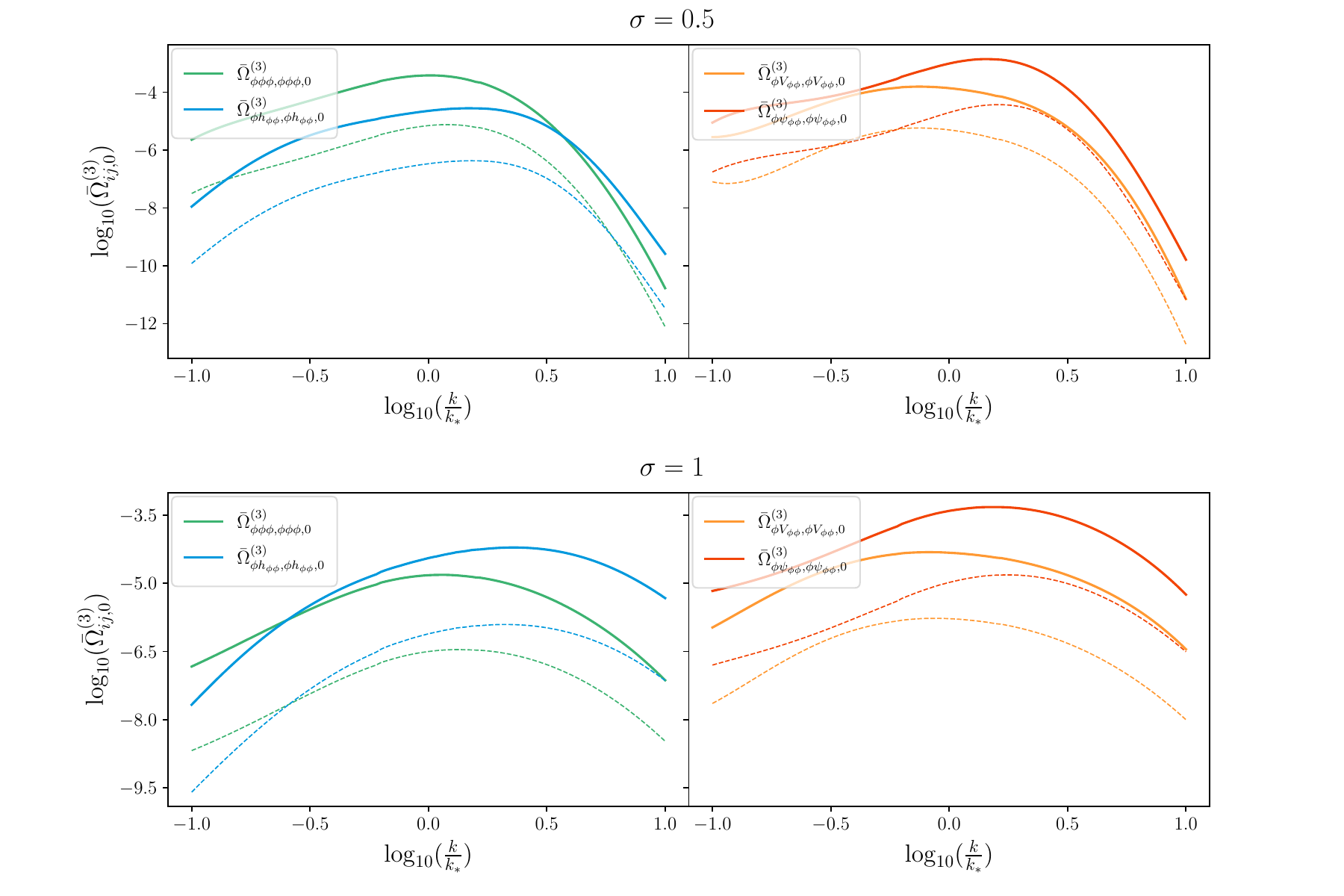}
    \caption{The energy density spectra of third order \acp{SIGW} for 
different kinds of source terms: $\bar{\Omega}^{(3)}_{ij,0}(k)$ with $i=j$. We set $A_\zeta = 1$ and $\sigma = 0.5, 1$. The solid lines represent the result of the power spectrum and the corresponding dashed lines represent the error in the calculation of the corresponding power spectrum.}\label{fig:33_diagonal_term}
\end{figure}
In this paper, we extend the results of Refs.~\cite{Zhou:2021vcw} to the log-normal primordial power spectrum
\begin{equation}
    \mathcal{P}_{\zeta}(k) = \frac{A_\zeta}{\sqrt{2\pi\sigma^2}}\exp\left(-\frac{\ln^{2}(k/k_*)}{2\sigma^2}\right) \ ,
\end{equation}
where $A_{\zeta}$ is the amplitude of primordial power spectrum and $k_*=2\pi f_*$ is the wavenumber at which the primordial power spectrum has a log-normal peak. Following Eq.~(\ref{eq:P3}), we numerically calculate the energy density spectra of third order \acp{SIGW} using \texttt{vegas} package \cite{Lepage:1977sw,Lepage:2020tgj,peter_lepage_2023_10402580}. The numerical results of third order energy density spectra $\bar{\Omega}^{(3)}_{ij,0}(k)$ with $i = j$ are shown in Fig.~\ref{fig:33_diagonal_term}. Since \texttt{vegas} uses a Monte Carlo integration strategy, it is necessary to analyze the errors in our calculations. As depicted in Fig.~\ref{fig:33_diagonal_term}, we present the computed results for the third order energy density spectrum using solid lines, while the corresponding numerical calculation errors are represented by dashed lines of the same color. Notably, the numerical integration error is negligible compared to the computed values. Moreover, in the case of a monochromatic primordial power spectrum, the second order vector perturbation induced by the first order scalar perturbation is zero \cite{Chang:2022dhh}. This implies that source term $\phi V_{\phi\phi}$ does not contribute to the third order \acp{SIGW} \cite{Zhou:2021vcw}. However, as shown by the orange solid line in Fig.~\ref{fig:33_diagonal_term}, considering a log-normal primordial power spectrum, we find that non-zero second order vector perturbations can impact the third order \acp{SIGW}.

\begin{figure}
    \centering
    \includegraphics[scale=0.7]{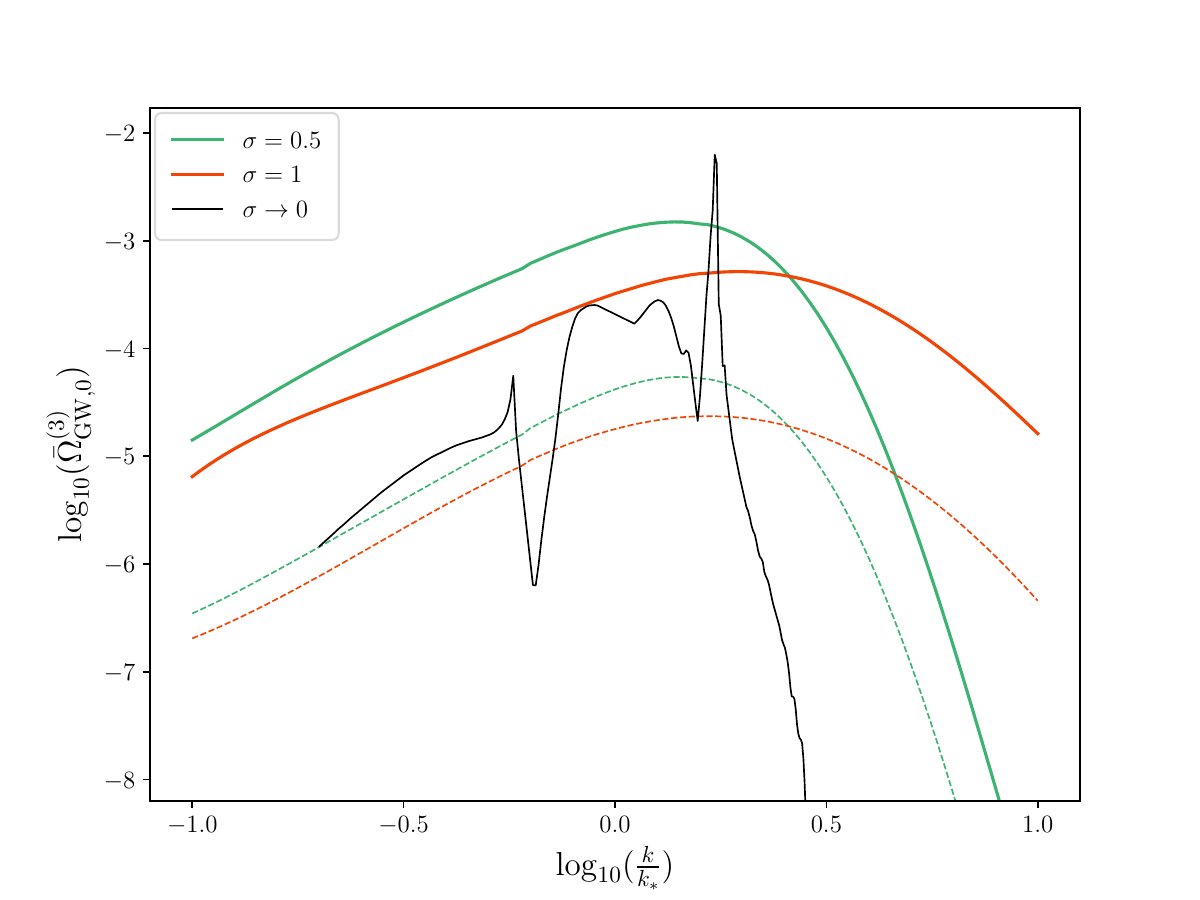}
    \caption{The energy density spectra of third order \acp{SIGW} $\bar{\Omega}_{\mathrm{GW},0}^{(3)}(k)$. The black line represents the result of monochromatic primordial power spectrum in Ref.~\cite{Zhou:2021vcw}. The green and red lines represent the cases of log-normal primordial power spectra with $\sigma=0.5$ and $\sigma=1$, respectively. The solid line is the result of the power spectrum and the dashed line of the same color indicates the error in the calculation of the corresponding power spectrum.}\label{fig:33_all}.
\end{figure}

\begin{figure}
    \centering
    \includegraphics[scale=0.5]{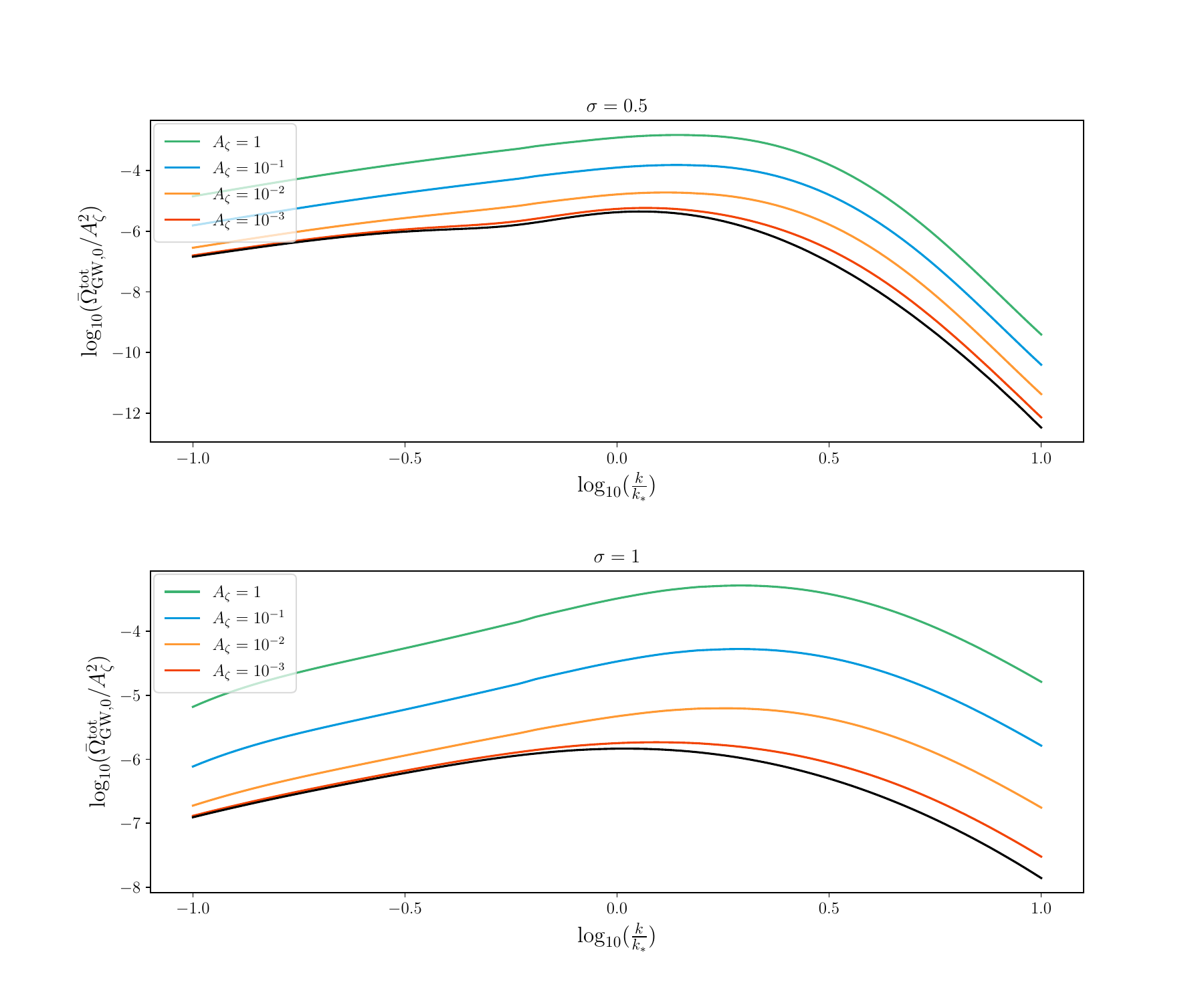}
    \caption{The total energy density spectra of \acp{SIGW} $\bar{\Omega}^{\mathrm{tot}}_{\mathrm{GW},0}(k)/A_{\zeta}^2$ for different values of $A_{\zeta}$. The black line represents the energy density of second order \acp{SIGW} $\bar{\Omega}^{(2)}_{\mathrm{GW},0}(k)/A_{\zeta}^2$.}\label{fig:tot}
\end{figure}

The energy density spectra of third order \acp{SIGW} $\bar{\Omega}_{\mathrm{GW},0}^{(3)}(k)$ for different $\sigma$ are shown in Fig.~\ref{fig:33_all}. As $\sigma$ gradually increases, the energy density spectrum of third-order gravitational waves widens progressively.
In addition, the total power spectrum of the second order and third order \acp{SIGW} $\bar{\Omega}^{\mathrm{tot}}_{\mathrm{GW},0}(k)$ $\left(=\bar{\Omega}_{\mathrm{GW},0}^{(2)}(k)+\bar{\Omega}_{\mathrm{GW},0}^{(3)}(k)\right)$ is plotted in Fig.\ref{fig:tot}. As depicted, even for $A_\zeta=10^{-2}$, the third order \acp{SIGW} still impact a noticeable correction to the total power spectrum. When $A_{\zeta}=10^{-3}$, the impact of third order \acp{SIGW} diminishes. Furthermore, the results presented in  Fig.\ref{fig:tot} indicate that the substantial contribution of third order \acp{SIGW} to the total energy density spectrum in the high-frequency region $(\tilde{k}=k/k_*>0.1)$ is not solely due to the peculiar nature of the monochromatic primordial power spectrum. Even when considering a more realistic broadened primordial power spectrum, third order \acp{SIGW} continue to exert a significant impact on the total energy density spectrum.

\section{Detection of \acp{SIGW}}\label{sec:4.0}
In this section, we analyze the \ac{SNR} of \ac{LISA} and also apply our results to a Bayesian analysis of the \ac{PTA} data in the case where \acp{SIGW} dominates \acp{SGWB} in the \ac{LISA} and \ac{PTA} bands respectively.

\subsection{Signal-to-noise ratio of LISA}\label{sec:4.1}
\begin{figure*}[htbp]
    \captionsetup{
      justification=raggedright,
      singlelinecheck=true
    }
    \centering

	\subfloat[$\sigma = 0.5$ ]{\includegraphics[width=.5\columnwidth]{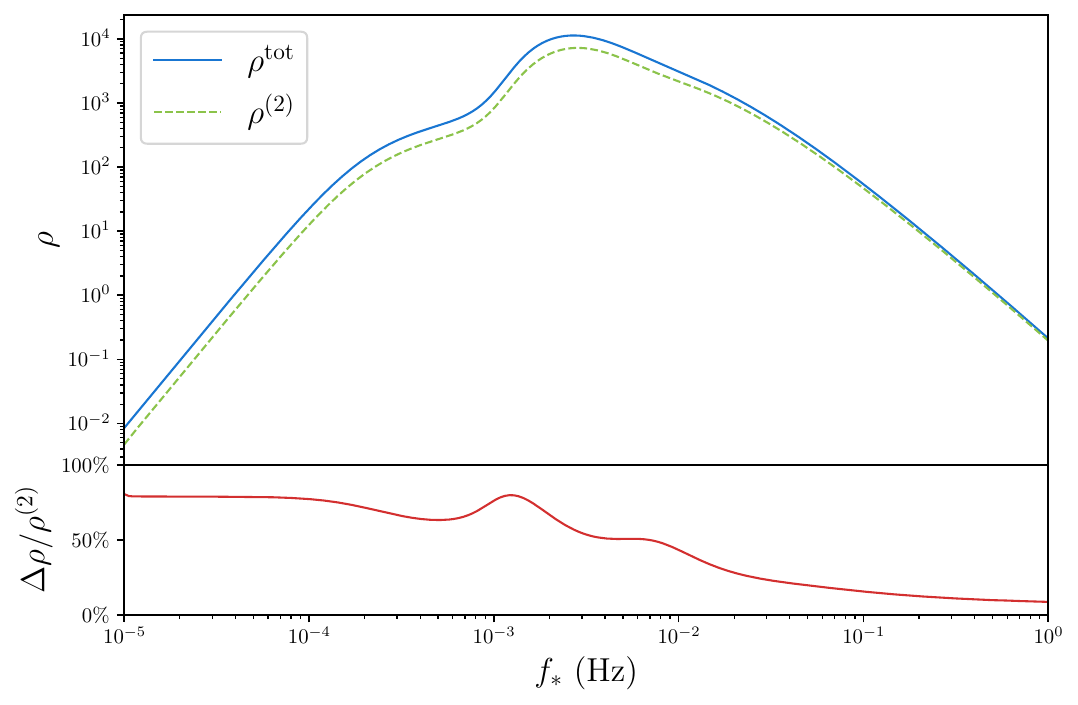}}
        \subfloat[$\sigma = 1$ ]{\includegraphics[width=.5\columnwidth]{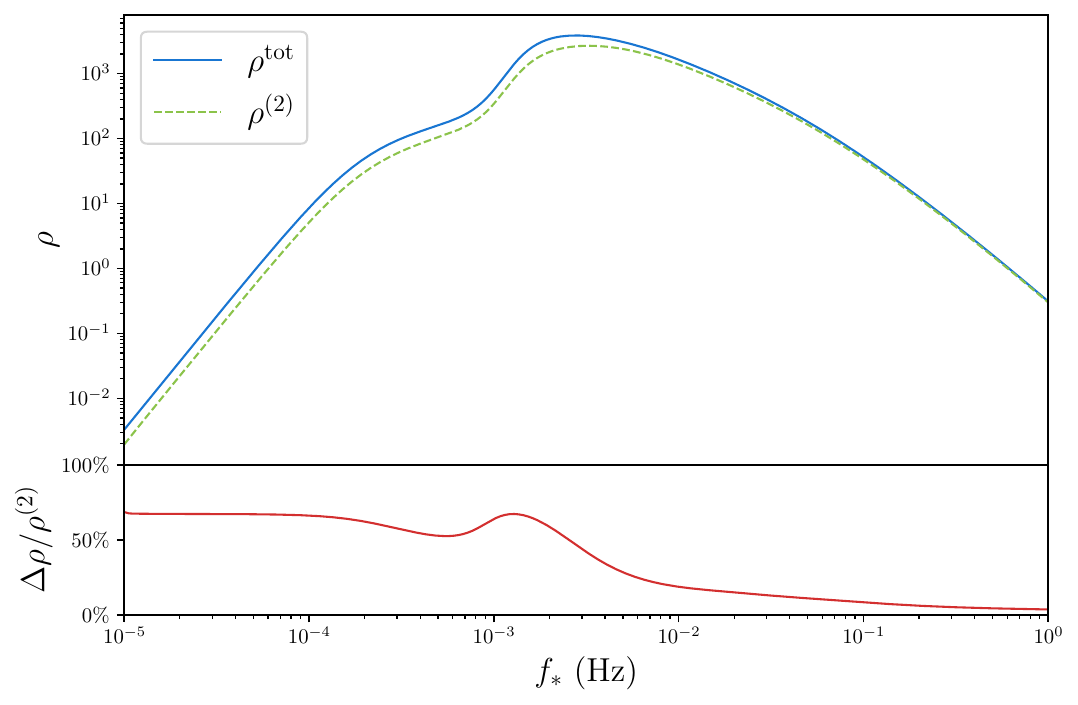}}
        
\caption{\label{fig:LISA_fixA} The \ac{SNR} as a function of frequency for LISA experiment for both the second order \acp{SIGW} $\rho^{(2)}$ and the total \acp{SIGW} $\rho^{\mathrm{tot}} = \rho^{(2)} + \rho^{(3)}$ with a fixed value of $A_\zeta = 0.02$. The curves below show the improvement of \ac{SNR} after considering the third order \acp{SIGW}.}
\end{figure*}

\begin{figure*}[htbp]
    \captionsetup{
      justification=raggedright,
      singlelinecheck=true
    }
    \centering

	\subfloat[$\sigma = 0.5$ ]{\includegraphics[width=.5\columnwidth]{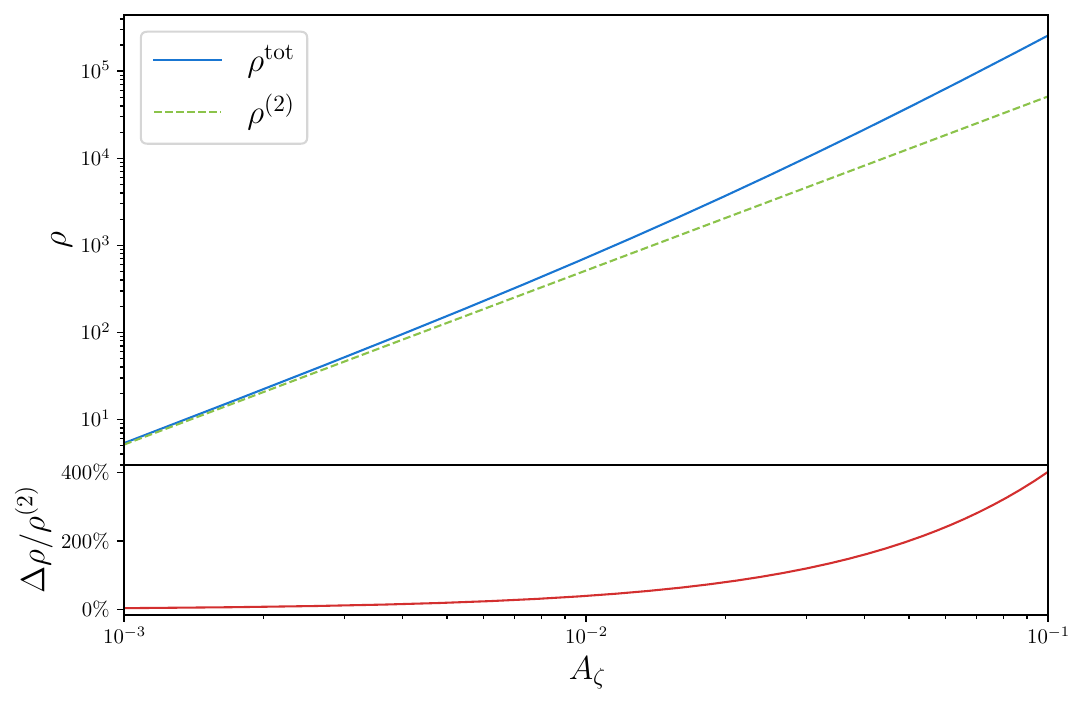}}
        \subfloat[$\sigma = 1$ ]{\includegraphics[width=.5\columnwidth]{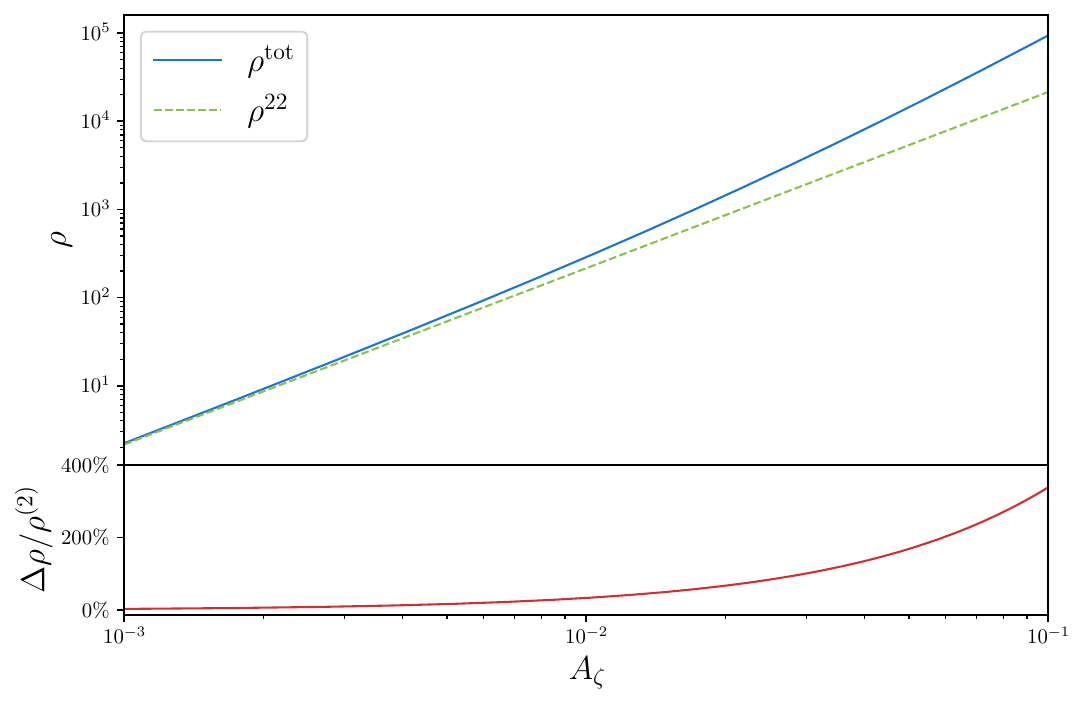}}
        
\caption{\label{fig:LISA_fixf} The \ac{SNR} as a function of frequency for LISA experiment for both the second order \acp{SIGW} $\rho^{(2)}$ and the total \acp{SIGW} $\rho^{\mathrm{tot}} = \rho^{(2)} + \rho^{(3)}$ with a fixed value of $f_* = 1.3 \times 10^{-3} \mathrm{Hz}$. The curves below show the improvement of \ac{SNR} after considering the third order \acp{SIGW}.}
\end{figure*}

In order to explore the effect of third order \acp{SIGW} on high-frequency region $(\tilde{k}=k/k_*>0.1)$, we calculate the \ac{SNR} $\rho$ for \ac{LISA} \cite{Siemens:2013zla, Robson:2018ifk, Zhao:2022kvz}
\begin{equation}
    \rho = \sqrt{T}\left[ \int \mathrm{d} f\left(\frac{\bar{\Omega}_{\mathrm{GW},0}(f)}{\Omega_\mathrm{n}(f)}\right)^2\right]^{1/2} \ ,
\end{equation}
where $T$ is the observation time and we set $T=4$ years here. $\Omega_\mathrm{n}(f)=2\pi^2f^3S_n/3H_0^2$, where $H_0$ is the Hubble constant, $S_n$ is the strain noise power spectral density \cite{Robson:2018ifk}. In Fig.~\ref{fig:LISA_fixA}, we plot the \ac{SNR} curves for LISA experiments. $\rho^{(2)}$ and $\rho^{\mathrm{tot}} = \rho^{(2)} + \rho^{(3)}$ represent the \ac{SNR} for the second order contribution and total contribution, respectively. It shows that the third order \acp{SIGW} have a large impact in high-frequency region ($\tilde{k}>0.1$) and can produce a $70\%$ \ac{SNR} boost when $A_\zeta = 0.02$. Fig.~\ref{fig:LISA_fixf} shows the variation of the \ac{SNR} with $A_\zeta$ for a fixed $f_*$. If we tolerate a value of $A_\zeta$ as high as $A_{\zeta}\sim 0.1$, then the \ac{SNR} can even be improved by $400\%$ at $f_* = 1.3\times10^{-3}\mathrm{Hz}$. Therefore, the substantial contribution of third order \acp{SIGW} in high-frequency region ($\tilde{k}>0.1$) is not solely a consequence of the special nature of monochromatic primordial power spectrum. Their impact remains significant even when considering realistic power spectra.

\subsection{PTA observations}\label{sec:4.2}
\begin{figure*}[htbp]
    \captionsetup{
      justification=raggedright,
      singlelinecheck=true
    }
    \centering

	\subfloat[$\sigma = 0.5$ with NANOGrav 15-year only]{\includegraphics[width=.4\columnwidth]{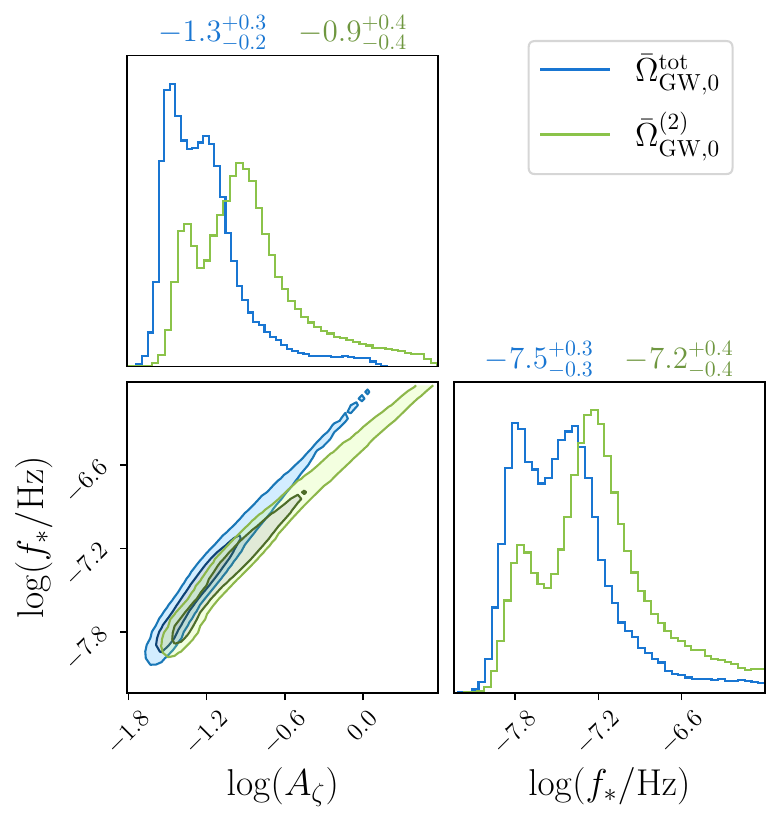}}
        \subfloat[$\sigma = 0.5$ with the combined data]{\includegraphics[width=.4\columnwidth]{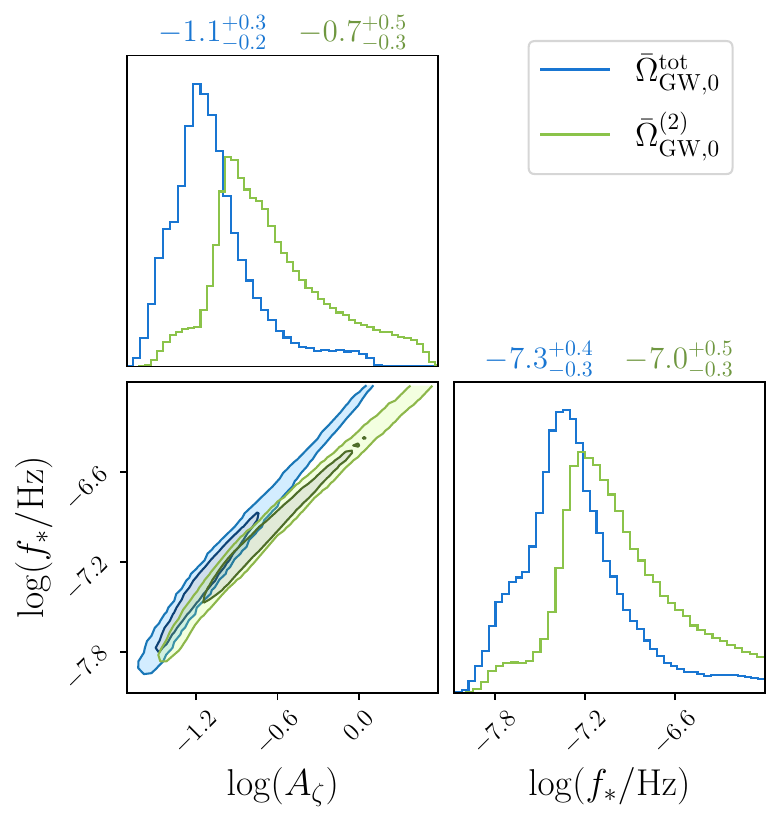}}\\
        \subfloat[$\sigma = 1$ with NANOGrav 15-year only]{\includegraphics[width=.4\columnwidth]{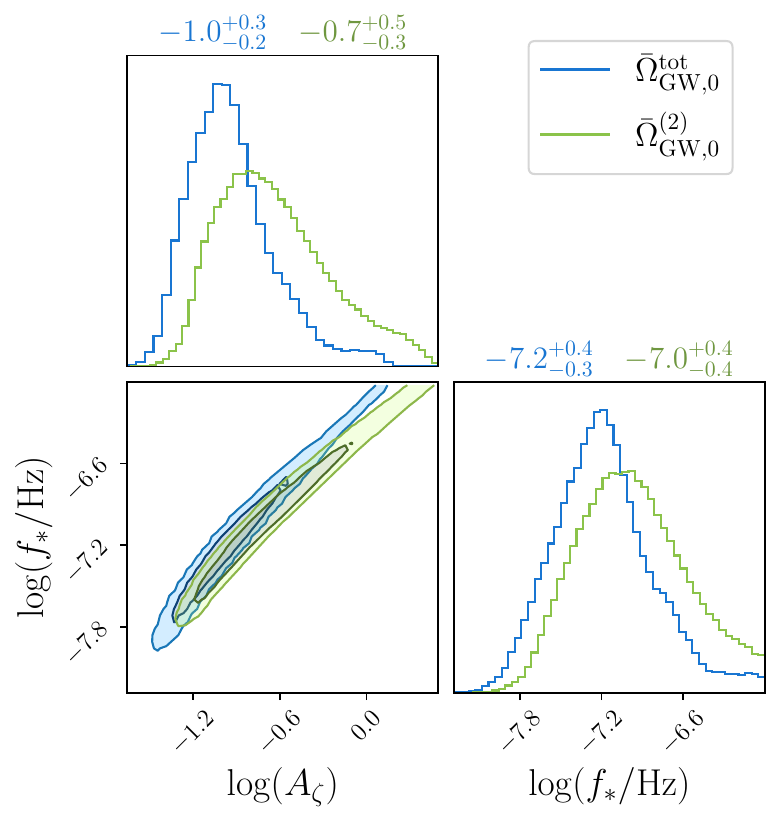}}
        \subfloat[$\sigma = 1$ with the combined data]{\includegraphics[width=.4\columnwidth]{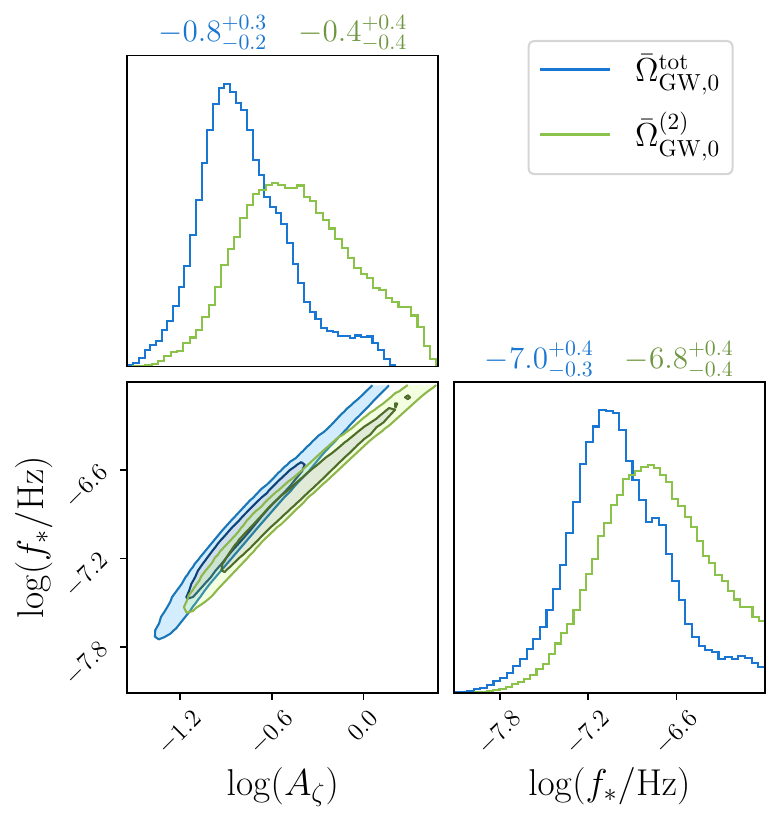}} 
        
\caption{\label{fig:PTA} The posteriors and $1$-$\sigma$ ranges of two independent parameters $\log(f_*/\mathrm{Hz})$ and $\log(A_\zeta)$. The blue one corresponds to the result of $\bar{\Omega}^{\mathrm{tot}}_{\mathrm{GW},0}(k)\left (=\bar{\Omega}^{(2)}_{\mathrm{GW},0}(k)+\bar{\Omega}^{(3)}_{\mathrm{GW},0}(k)\right)$ and the green one corresponds to the result of $\bar{\Omega}^{(2)}_{\mathrm{GW},0}(k)$ only. The term 'combined data' refers to the aggregate dataset encompassing both NANOGrav 15-year and EPTA DR2 new. The numbers above the figures correspond to the median values and $1$-$\sigma$ ranges of the parameters.}
\end{figure*}

\begin{figure}[htbp]
    \centering
    \includegraphics[width=.95\columnwidth]{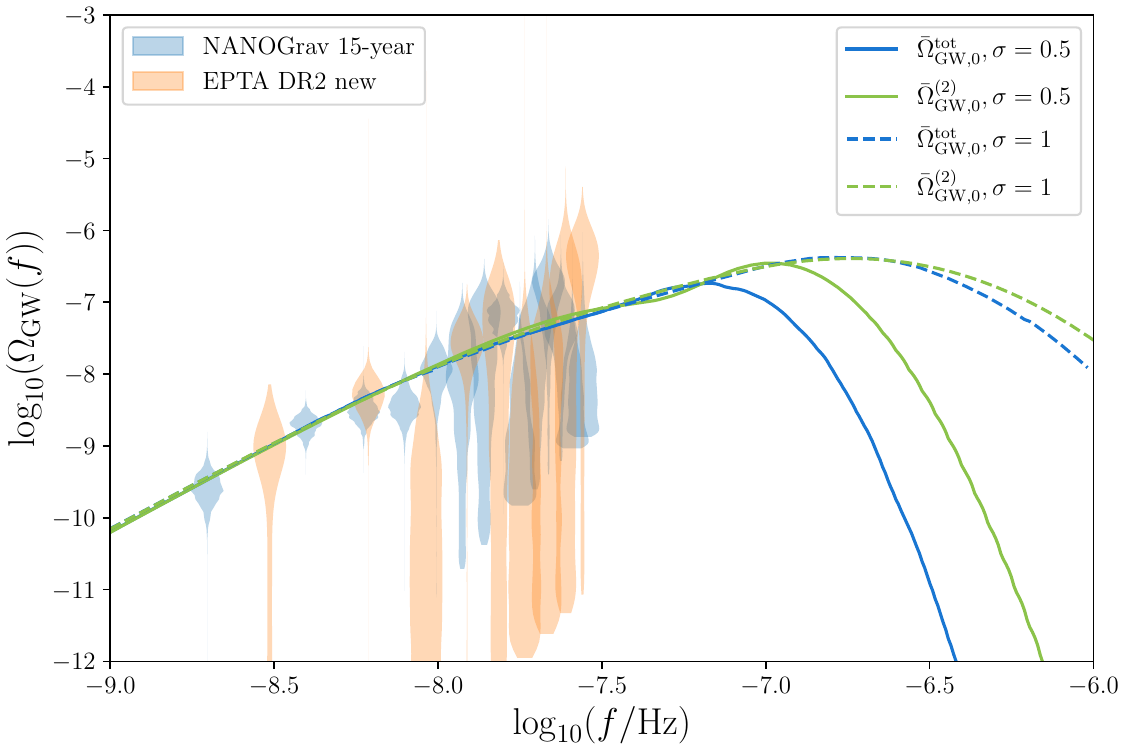}
\caption{\label{fig:violin} The energy density spectra with $\sigma=1, 0.5$. The blue curves correspond to the total density spectra $\bar{\Omega}^{\mathrm{tot}}_{\mathrm{GW},0}(k)\left (=\bar{\Omega}^{(2)}_{\mathrm{GW},0}(k)+\bar{\Omega}^{(3)}_{\mathrm{GW},0}(k)\right)$ and the green curves correspond to the density spectra of $\bar{\Omega}^{(2)}_{\mathrm{GW},0}(k)$ only. We select the parameters from subfigures (b) and (d) of Fig. \ref{fig:PTA}. The energy density spectra derived from the free spectrum of NANOGrav 15-year data set and EPTA DR2 new are also shown here.}
\end{figure}

In order to constrain the parameter space of primordial power spectrum in terms of PTA observations, we use Ceffyl \cite{lamb2023rapid} package embedded in PTArade \cite{mitridate2023ptarcade} to analyze the data from the first 14 frequency bins of NANOGrav 15-year and the first 9 frequency bins in EPTA DR2 new. We show the posteriors distributions in Fig.~\ref{fig:PTA}, where the priors distributions of $\log(f_*/\mathrm{Hz})$ and $\log(A_\zeta)$ are set as uniform distributions over the intervals $[-9,-6]$ and $[-3,1]$, respectively. Analyzing the posterior distributions of second order \acp{SIGW} energy density spectrum $\bar{\Omega}^{(2)}_{\mathrm{GW},0}(k)$ as well as the total energy density spectrum $\bar{\Omega}^{\mathrm{tot}}_{\mathrm{GW},0}(k)$ $\left(=\bar{\Omega}^{(2)}_{\mathrm{GW},0}(k)+\bar{\Omega}^{(3)}_{\mathrm{GW},0}(k)\right)$, we find that, in low-frequency regime ($\tilde{k}<0.1$), the effect of the third order \acp{SIGW} is smaller than in high-frequency region. The median values of $\log(A_\zeta)$ and $\log(f_*/\mathrm{Hz})$ become smaller when considering the effects of third order \acp{SIGW}. More precisely,
\begin{enumerate}[left=1cm, rightmargin=1cm]
  \item For $\sigma = 0.5$, in the case of NANOGrav 15-year data only, the median value of $\log(A_\zeta)$ goes from $-0.9$ to $-1.3$ and the median value of $\log(f_*/\mathrm{Hz})$ goes from $-7.2$ to $-7.5$. In the case of joint data from NANOGrav 15-year and EPTA DR2 new, the median value of $\log(A_\zeta)$ goes from $-0.7$ to $-1.1$ and the median value of $\log(f_*/\mathrm{Hz})$ goes from $-7.0$ to $-7.3$.
  \item For $\sigma = 1$, in the case of NANOGrav 15-year data only, the median value of $\log(A_\zeta)$ goes from $-0.7$ to $-1.0$ and the median value of $\log(f_*/\mathrm{Hz})$ goes from $-7.0$ to $-7.2$. In the case of joint data from NANOGrav 15-year and EPTA DR2 new, the median value of $\log(A_\zeta)$ goes from $-0.4$ to $-0.8$ and the median value of $\log(f_*/\mathrm{Hz})$ goes from $-6.8$ to $-7.0$.
\end{enumerate}
In Fig.~\ref{fig:violin}, we plot the total energy density spectra $\bar{\Omega}^{\mathrm{tot}}_{\mathrm{GW,0}}(k)$ and compare the results with the case of second order \acp{SIGW}.

\section{Conclusion and discussion}\label{sec:5.0}
In this paper, we systematically investigated the third order \acp{SIGW} with log-normal primordial power spectra. Our results reveal that the third order \acp{SIGW} have a significant impact on the total energy density spectrum of \acp{GW}. More precisely, their contribution is most pronounced in the high-frequency region of the energy density spectrum. As depicted in Fig.\ref{fig:tot}, the presence of third order \acp{SIGW} alters the shape of the energy density spectrum at high frequencies. Quantifying their influence, we computed the \ac{SNR} for the \ac{LISA} frequency band. Remarkably, when the amplitude of the primordial power spectrum $A_{\zeta}=0.02$, third order \acp{SIGW} enhance the \ac{SNR} of \ac{LISA} by approximately $70\%$. However, in the low-frequency regime of the total gravitational wave energy density spectrum, their contribution is relatively small, as shown in Fig.~\ref{fig:PTA}. Combining this with observational data from \ac{PTA}, we find that third order \acp{SIGW} also impact the amplitude $A_{\zeta}$ and peak position $f_*$ of the log-normal primordial power spectrum. In summary, we conclude that the substantial contribution of third order \acp{SIGW} is not solely a consequence of the special nature of monochromatic primordial power spectrum. Their impact remains significant even when considering realistic primordial power spectra.

The energy density spectrum of third order \acp{SIGW} can be obtained by calculating the two-point correlation function of third order \acp{SIGW}. As shown in Appendix.~\ref{sec:C}, the two-point correlation function of third order \acp{SIGW} corresponds to the two-loop diagrams proportional to $A_{\zeta}^3$ in cosmological perturbation theory. It is important to note that the two-point correlation function of fourth order \acp{SIGW} and second order \acp{SIGW} $\langle h^{\lambda,(4)}_{\mathbf{k}} h^{\lambda ',(2)}_{\mathbf{k}'}\rangle$ can still provide a two-loop contribution proportional to $A_{\zeta}^3$. Since the contribution of the two-point correlation function $\langle h^{\lambda,(4)}_{\mathbf{k}} h^{\lambda ',(2)}_{\mathbf{k}'}\rangle$ is also proportional to $A_{\zeta}^3$, it may also have a significant contribution to the total energy density spectrum of \acp{GW}. However, due to the complexity of the calculation of fourth order \acp{SIGW}, the calculations of the two-loop contributions of \acp{SIGW} waves have not yet been completed.

We can observe from the above discussion that the third order \acp{SIGW} have a significant impact on the total energy density spectrum of \acp{SIGW}. As $A_\zeta$ increases, the contribution of third order \acp{SIGW} can even surpass that of second order ones. This makes it necessary to consider the convergence of the perturbative expansion in the calculation of \acp{SIGW}. The total energy density spectrum of \acp{SIGW} can be expressed in the following form
\begin{equation}\label{eq:WRZK}
\begin{aligned}
\Omega_{\mathrm{GW}}(k)&=A_\zeta^2\left(\Omega_{\mathrm{GW}}^{(2,2)}(k)\right)\bigg|_{A_\zeta=1}+A_\zeta^3\left(\sum^{1}_{m=0}~\Omega_{\mathrm{GW}}^{(3+m,3-m)}(k)\right)\bigg|_{A_\zeta=1} \\
&+A_\zeta^4\left(\sum^{2}_{m=0}~\Omega_{\mathrm{GW}}^{(4+m,4-m)}(k)\right)\bigg|_{A_\zeta=1}+\cdots \ , 
\end{aligned}
\end{equation}
where, $\left(\Omega_{\mathrm{GW}}^{(n,m)}(k)\right)\bigg|_{A_\zeta=1}$ denotes the energy density spectrum corresponds to the two-point correlation function $\langle h^{\lambda,(n)}_{\mathbf{k}} h^{\lambda ',(m)}_{\mathbf{k}'}\rangle$ when the amplitude of the primordial power spectrum $A_\zeta = 1$. We have denoted $\Omega_{\mathrm{GW}}^{(n,n)}(k)$ as $\Omega_{\mathrm{GW}}^{(n)}(k)$ in previous sections. Similar to second and third order \acp{SIGW}, the energy density spectrum $\Omega_{\mathrm{GW}}^{(n,m)}(k)$ can be approximated as
\begin{equation}\label{eq:dis}
\begin{aligned}
\Omega_{\mathrm{GW}}^{(n,m)}(k)\sim \int & d^3 p_{n-1} \cdot\cdot\cdot\int d^3 p_1\mathbb{P}^{ij} ~ I^{(n)} I^{(m)} \\
&\times \mathcal{P}_{\zeta}\left(| \mathbf{k}-\mathbf{p}_1|\right)\cdots\mathcal{P}_{\zeta}\left(|\mathbf{p}_{n-2} - \mathbf{p}_{n-1}| \right)\mathcal{P}_{\zeta}\left(| \mathbf{p}_{n-1}| \right) \ ,
\end{aligned}
\end{equation}
where, $I^{(n)}$ represents the kernel function of $n$-th order \acp{SIGW},  $\mathbb{P}^{ij}$ denotes the momentum polynomial, and $\mathcal{P}_{\zeta}\left(k \right)$ stands for the primordial power spectrum. Since the observational data of \ac{PTA} suggests that $A_\zeta$ has a relatively large value, in this case, third order \acp{SIGW} will have a more significant contribution compared to second order ones. This motivates us to consider the following three questions: 

1. Whether the two-point correlation function $\langle h^{\lambda,(4)}_{\mathbf{k}} h^{\lambda ',(2)}_{\mathbf{k}'}\rangle\sim A^{(3)}_{\zeta}$ and higher order \acp{SIGW} would generate contributions larger than the second and third order \acp{SIGW}?

2. How the two-point correlation function $\langle h^{\lambda,(4)}_{\mathbf{k}} h^{\lambda ',(2)}_{\mathbf{k}'}\rangle\sim A^{(3)}_{\zeta}$ and higher order \acp{SIGW} affect the \acp{SNR} of LISA and the posteriors distributions of $\log(f_*/\mathrm{Hz})$ and $\log(A_\zeta)$ in PTA observation?

3. Under what conditions does the perturbative expansion of the \acp{SIGW} in Eq.~(\ref{eq:WRZK}) converge? 

However, without explicit calculations of the energy density spectra of higher order \acp{SIGW}, it is difficult to provide precise answers to the above three questions. Detailed calculation of high order \acp{SIGW} and convergence of the expansion in Eq.~(\ref{eq:WRZK}) requires further research in future work.

In the studies of \acp{SIGW}, the \acp{GW} are induced by large primordial curvature perturbations on small scales. For primordial tensor perturbations, there are not many observational constraints on small scales either. We can also consider models of large primordial tensor perturbations on small scales \cite{Gorji:2023sil}. In this case, the primordial curvature perturbation and the primordial tensor perturbation together induce high order \acp{GW}. Refs.~\cite{Chang:2022vlv,Chen:2022dah} studied the one-loop contribution of induced \acp{GW} in this case. The corresponding two-loop calculation and its constraints on specific inflationary models might be studied in future work.

\acknowledgments
We thank Dr. Xukun Zhang for useful discussions. This work has been funded by the National Nature Science Foundation of China under grant No. 12075249, 11690022, and 12275276, and the Key Research Program of the Chinese Academy of Sciences under Grant No.
XDPB15. 

\appendix
\section{Polarization tensors}\label{sec:A}
The polarization tensors for momentum $\mathbf{k}$ are defined as
\begin{equation}\label{eq:pt1}
	\begin{aligned}
		\varepsilon^{\times}_{ij}\left(\mathbf{k}  \right)=\frac{1}{\sqrt{2}}\left( e_i\left( \mathbf{k} \right)\bar{e}_j\left( \mathbf{k} \right)+\bar{e}_i\left( \mathbf{k} \right)e_j\left( \mathbf{k} \right)  \right) \ ,
	\end{aligned} 
\end{equation}
\begin{equation}
	\begin{aligned}
		\varepsilon^{+}_{ij}\left(\mathbf{k}  \right)=\frac{1}{\sqrt{2}}\left( e_i\left( \mathbf{k} \right)e_j\left( \mathbf{k} \right)-\bar{e}_i\left( \mathbf{k} \right)\bar{e}_j\left( \mathbf{k} \right)  \right) \ ,
	\end{aligned} 
\end{equation}
where $e_i\left( \mathbf{k} \right)$ and $\bar{e}_i\left( \mathbf{k} \right)$ are transverse polarization vectors in three dimensional momentum space. For a given coordinate system, the polarization vectors can be written as
\begin{equation}
	\begin{aligned}
		\mathbf{k}=\left(0,0,k  \right) \ , \ e_i\left( \mathbf{k} \right)=\left( 1,0,0 \right) \ , \ \bar{e}_i\left( \mathbf{k} \right)=\left( 0,1,0 \right) \ .
	\end{aligned} 
\end{equation}
In this coordinate system, the explicit expressions of polarization tensors $\varepsilon_{i j}^{\lambda}(\mathbf{k})$ $(\lambda=+,\times)$ can be obtained in terms of Eq.~(\ref{eq:pt1}). We define $|\mathbf{k}-\mathbf{p}|=u|\mathbf{k}|$ and $|\mathbf{p}|=v|\mathbf{k}|$. Then, the polarization tensors for $\mathbf{k}-\mathbf{p}$ and $\mathbf{p}$ are given by
\begin{eqnarray}\label{eq:pt2}
		\varepsilon^{\times}_{ij}\left(\mathbf{k}-\mathbf{p}  \right)&=&\frac{1}{\sqrt{2}}\left( e_i\left( \mathbf{k}-\mathbf{p} \right)\bar{e}_j\left( \mathbf{k}-\mathbf{p} \right)+\bar{e}_i\left( \mathbf{k}-\mathbf{p} \right)e_j\left( \mathbf{k}-\mathbf{p} \right)  \right) \ , \\
		\varepsilon^{+}_{ij}\left(\mathbf{k}-\mathbf{p}  \right)&=&\frac{1}{\sqrt{2}}\left( e_i\left( \mathbf{k}-\mathbf{p} \right)e_j\left( \mathbf{k}-\mathbf{p} \right)-\bar{e}_i\left( \mathbf{k}-\mathbf{p} \right)\bar{e}_j\left( \mathbf{k}-\mathbf{p} \right)  \right) \ , \\
		\varepsilon^{\times}_{ij}\left(\mathbf{p}  \right)&=&\frac{1}{\sqrt{2}}\left( e_i\left( \mathbf{p} \right)\bar{e}_j\left( \mathbf{p} \right)+\bar{e}_i\left( \mathbf{p} \right)e_j\left( \mathbf{p} \right)  \right)\ , \\ \varepsilon^{+}_{ij}\left(\mathbf{p}  \right)&=&\frac{1}{\sqrt{2}}\left( e_i\left( \mathbf{p} \right)e_j\left( \mathbf{p} \right)-\bar{e}_i\left( \mathbf{p} \right)\bar{e}_j\left( \mathbf{p} \right)  \right) \ ,
\end{eqnarray}
 where
\begin{eqnarray}
		\mathbf{k}-\mathbf{p}&=&k\left(-\sqrt{ v^2-\frac{1}{4}  \left(-u^2+v^2+1\right)^2},0,\frac{1}{2} \left(u^2-v^2+1\right)  \right)  \ , \\
		e_i\left( \mathbf{k}-\mathbf{p} \right)&=&\left(\frac{u^2-v^2+1}{2 u},0,\frac{\sqrt{-u^4+2 u^2 v^2+2 u^2-v^4+2 v^2-1}}{2 u}  \right)\ , \\
		\bar{e}_i\left( \mathbf{k}-\mathbf{p} \right)&=&\left( 0,1,0 \right) \ , \\
		\mathbf{p}&=&k\left(\sqrt{ v^2-\frac{1}{4} \left(-u^2+v^2+1\right)^2},0,\frac{1}{2}  \left(-u^2+v^2+1\right) \right)  \ , \\
		e_i\left( \mathbf{p} \right)&=&\left(-\frac{-u^2+v^2+1}{2 v},0,\frac{\sqrt{-u^4+2 u^2 \left(v^2+1\right)-\left(v^2-1\right)^2}}{2 v}  \right)\ , \\
		\bar{e}_i\left( \mathbf{p} \right)&=&\left( 0,1,0 \right) \ .
\end{eqnarray}

\section{Six-point function}\label{sec:B}
The non-trivial terms in Wick’s expansions of six-point correlation function are 
\begin{equation}\label{eq:Wick}
    \begin{aligned}
& \left\langle\zeta_{\mathbf{k}-\mathbf{p}} \zeta_{\mathbf{p}-\mathbf{q}} \zeta_{\mathbf{q}} \zeta_{\mathbf{k}^{\prime}-\mathbf{p}^{\prime}} \zeta_{\mathbf{p}^{\prime}-\mathbf{q}^{\prime}} \zeta_{\mathbf{q}^{\prime}}\right\rangle \\
&=\langle \zeta_{\mathbf{k}-\mathbf{p}}\zeta_{\mathbf{k}'-\mathbf{p}'} \rangle \langle\zeta_{\mathbf{p}-\mathbf{q}}\zeta_{\mathbf{p}'-\mathbf{q}'}\rangle \langle\zeta_{\mathbf{q}}\zeta_{\mathbf{q}'}\rangle +\langle \zeta_{\mathbf{k}-\mathbf{p}}\zeta_{\mathbf{k}'-\mathbf{p}'} \rangle \langle\zeta_{\mathbf{p}-\mathbf{q}}\zeta_{\mathbf{q}'}\rangle \langle\zeta_{\mathbf{q}}\zeta_{\mathbf{p}'-\mathbf{q}'}\rangle \\
&+\langle \zeta_{\mathbf{k}-\mathbf{p}} \zeta_{\mathbf{p}'-\mathbf{q}'}\rangle \langle \zeta_{\mathbf{p}-\mathbf{q}} \zeta_{\mathbf{k}'-\mathbf{p}'} \rangle \langle \zeta_{\mathbf{q}} \zeta_{\mathbf{q}'} \rangle +\langle \zeta_{\mathbf{k}-\mathbf{p}} \zeta_{\mathbf{p}'-\mathbf{q}'}\rangle \langle\zeta_{\mathbf{p}-\mathbf{q}}\zeta_{\mathbf{q}'} \rangle \langle\zeta_{\mathbf{q}}\zeta_{\mathbf{k}'-\mathbf{p}'} \rangle \\
& +\langle \zeta_{\mathbf{k}-\mathbf{p}} \zeta_{\mathbf{q}'} \rangle \langle\zeta_{\mathbf{p}-\mathbf{q}} \zeta_{\mathbf{k}'-\mathbf{p}'}\rangle \langle\zeta_{\mathbf{q}} \zeta_{\mathbf{p}'-\mathbf{q}'}\rangle +\langle \zeta_{\mathbf{k}-\mathbf{p}} \zeta_{\mathbf{q}'} \rangle \langle\zeta_{\mathbf{p}-\mathbf{q}}\zeta_{\mathbf{p}'-\mathbf{q}'} \rangle \langle\zeta_{\mathbf{q}}\zeta_{\mathbf{k}'-\mathbf{p}'} \rangle\\
& =\frac{(2\pi^2)^3}{(k-p)^3(p-q)^3 q^3} \delta\left(\mathbf{k}+\mathbf{k}^{\prime}\right)\delta\left(\mathbf{p}+\mathbf{p}^{\prime}\right) \delta\left(\mathbf{q}+\mathbf{q}^{\prime}\right) P_{\zeta}(k-p) P_{\zeta}(p-q) P_{\zeta}(q) \\
& +\frac{(2\pi^2)^3}{(k-p)^3(p-q)^3 q^3} \delta\left(\mathbf{k}+\mathbf{k}^{\prime}\right)\delta\left(\mathbf{p}+\mathbf{p}^{\prime}\right) \delta\left(\mathbf{q}+\mathbf{p}^{\prime}-\mathbf{q}^{\prime}\right) P_{\zeta}(k-p) P_{\zeta}(p-q) P_{\zeta}(q) \\
& +\frac{(2\pi^2)^3}{(k-p)^3(p-q)^3 q^3} \delta\left(\mathbf{k}+\mathbf{k}^{\prime}\right)\delta\left(\mathbf{p}-\mathbf{q}-\mathbf{k}-\mathbf{p}^{\prime}\right) \delta\left(\mathbf{q}+\mathbf{q}^{\prime}\right) P_{\zeta}(k-p) P_{\zeta}(p-q) P_{\zeta}(q) \\
& +\frac{(2\pi^2)^3}{(k-p)^3(p-q)^3 q^3} \delta\left(\mathbf{k}+\mathbf{k}^{\prime}\right)\delta\left(\mathbf{p}-\mathbf{q}+\mathbf{q}^{\prime}\right) \delta\left(\mathbf{q}-\mathbf{k}-\mathbf{p}^{\prime}\right) P_{\zeta}(k-p) P_{\zeta}(p-q) P_{\zeta}(q) \\
& +\frac{(2\pi^2)^3}{(k-p)^3(p-q)^3 q^3} \delta\left(\mathbf{k}+\mathbf{k}^{\prime}\right)\delta\left(\mathbf{p}-\mathbf{k}-\mathbf{q}^{\prime}\right) \delta\left(\mathbf{q}+\mathbf{p}^{\prime}-\mathbf{q}^{\prime}\right) P_{\zeta}(k-p) P_{\zeta}(p-q) P_{\zeta}(q) \\
& +\frac{(2\pi^2)^3}{(k-p)^3(p-q)^3 q^3} \delta\left(\mathbf{k}+\mathbf{k}^{\prime}\right)\delta\left(-\mathbf{k}+\mathbf{p}-\mathbf{q}^{\prime}\right) \delta\left(\mathbf{q}-\mathbf{k}-\mathbf{p}^{\prime}\right) P_{\zeta}(k-p) P_{\zeta}(p-q) P_{\zeta}(q)\ .
\end{aligned}
\end{equation}

\section{Two-loop diagrams of third order \acp{SIGW}}\label{sec:C}
The two-loop diagrams of  of third order \acp{SIGW} are shown in following figures.

\begin{figure*}[htbp]
    \captionsetup{
      justification=raggedright,
      singlelinecheck=true
    }
    \centering

	\subfloat[$ \langle \zeta_{\mathbf{p-q}}\zeta_{\mathbf{p}'-\mathbf{q}'} \rangle \langle \zeta_{\mathbf{q}}\zeta_{\mathbf{q}'} \rangle \langle \zeta_{\mathbf{k}-\mathbf{p}}\zeta_{\mathbf{k}'-\mathbf{p}'} \rangle $]{\includegraphics[width=.33\columnwidth]{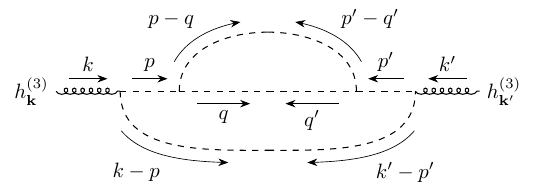}}
        \subfloat[$\langle \zeta_{\mathbf{k}-\mathbf{p}}\zeta_{\mathbf{q}'} \rangle  \langle \zeta_{\mathbf{p-q}}\zeta_{\mathbf{p}'-\mathbf{q}'} \rangle \langle \zeta_{\mathbf{q}}\zeta_{\mathbf{k}'-\mathbf{p}'} \rangle $]{\includegraphics[width=.33\columnwidth]{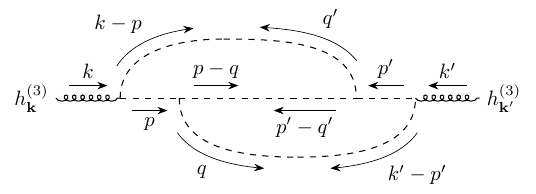}}
        \subfloat[$\langle \zeta_{\mathbf{k}-\mathbf{p}}\zeta_{\mathbf{q}'} \rangle \langle \zeta_{\mathbf{q}} \zeta_{\mathbf{p}'-\mathbf{q}'}\rangle \langle \zeta_{\mathbf{p-q}}\zeta_{\mathbf{k}'-\mathbf{p}'} \rangle$]{\includegraphics[width=.33\columnwidth]{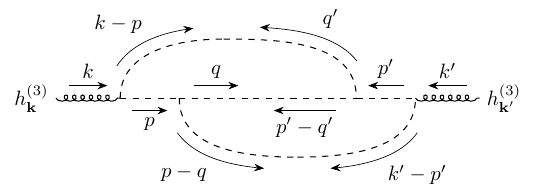}} \\

        \subfloat[$ \langle \zeta_{\mathbf{p-q}}\zeta_{\mathbf{q}'} \rangle \langle \zeta_{\mathbf{q}}\zeta_{\mathbf{p}'-\mathbf{q}'} \rangle \langle \zeta_{\mathbf{k}-\mathbf{p}}\zeta_{\mathbf{k}'-\mathbf{p}'} \rangle $]{\includegraphics[width=.33\columnwidth]{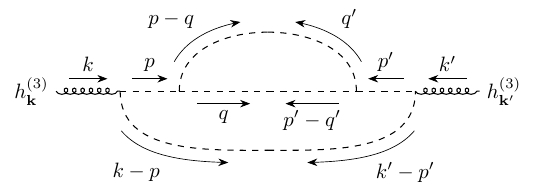}}
        \subfloat[$\langle \zeta_{\mathbf{k}-\mathbf{p}}\zeta_{\mathbf{p}'-\mathbf{q}'} \rangle \langle \zeta_{\mathbf{q}} \zeta_{\mathbf{q}'}\rangle \langle \zeta_{\mathbf{p-q}} \zeta_{\mathbf{k}'-\mathbf{p}'} \rangle$]{\includegraphics[width=.33\columnwidth]{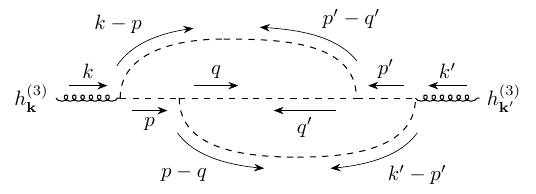}}
        \subfloat[$\langle \zeta_{\mathbf{k}-\mathbf{p}}\zeta_{\mathbf{p}'-\mathbf{q}'} \rangle  \langle \zeta_{\mathbf{p-q}}\zeta_{\mathbf{q}'} \rangle \langle \zeta_{\mathbf{q}} \zeta_{\mathbf{k}'-\mathbf{p}'}\rangle $]{\includegraphics[width=.33\columnwidth]{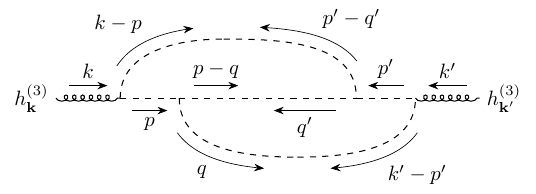}} 
        
\caption{\label{fig:FeynDiag33-1}The two-point function $\langle  h^{\lambda,(3)}_{\mathbf{k},\phi\psi_{\phi\phi}} h^{\lambda',(3)}_{\mathbf{k}',\phi\psi_{\phi\phi}}\rangle$. The diagrams correspond to the Wick’s expansions of six-point correlation function. The dashed lines, wavy lines and spring-like lines represent scalar, vector, and tensor perturbations, respectively.}
\end{figure*}

\begin{figure*}[htbp]
    \captionsetup{
      justification=raggedright,
      singlelinecheck=true
    }
    \centering

	\subfloat[$ \langle \zeta_{\mathbf{p-q}}\zeta_{\mathbf{p}'-\mathbf{q}'} \rangle \langle \zeta_{\mathbf{q}}\zeta_{\mathbf{q}'} \rangle \langle \zeta_{\mathbf{k}-\mathbf{p}}\zeta_{\mathbf{k}'-\mathbf{p}'} \rangle $]{\includegraphics[width=.33\columnwidth]{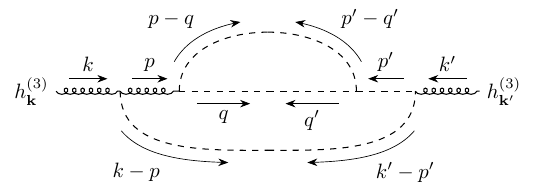}}
        \subfloat[$\langle \zeta_{\mathbf{k}-\mathbf{p}}\zeta_{\mathbf{q}'} \rangle  \langle \zeta_{\mathbf{p-q}}\zeta_{\mathbf{p}'-\mathbf{q}'} \rangle \langle \zeta_{\mathbf{q}}\zeta_{\mathbf{k}'-\mathbf{p}'} \rangle $]{\includegraphics[width=.33\columnwidth]{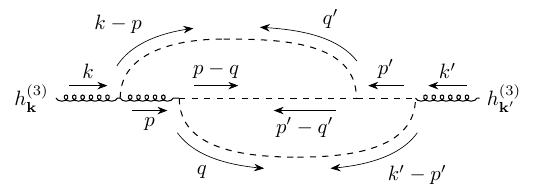}}
        \subfloat[$\langle \zeta_{\mathbf{k}-\mathbf{p}}\zeta_{\mathbf{q}'} \rangle \langle \zeta_{\mathbf{q}} \zeta_{\mathbf{p}'-\mathbf{q}'}\rangle \langle \zeta_{\mathbf{p-q}}\zeta_{\mathbf{k}'-\mathbf{p}'} \rangle$]{\includegraphics[width=.33\columnwidth]{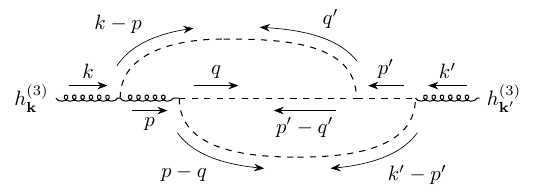}} \\

        \subfloat[$ \langle \zeta_{\mathbf{p-q}}\zeta_{\mathbf{q}'} \rangle \langle \zeta_{\mathbf{q}}\zeta_{\mathbf{p}'-\mathbf{q}'} \rangle \langle \zeta_{\mathbf{k}-\mathbf{p}}\zeta_{\mathbf{k}'-\mathbf{p}'} \rangle $]{\includegraphics[width=.33\columnwidth]{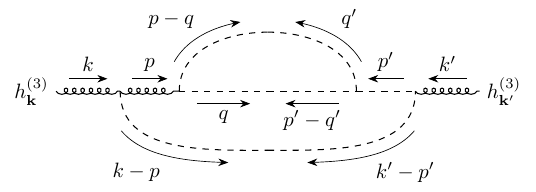}}
        \subfloat[$\langle \zeta_{\mathbf{k}-\mathbf{p}}\zeta_{\mathbf{p}'-\mathbf{q}'} \rangle \langle \zeta_{\mathbf{q}} \zeta_{\mathbf{q}'}\rangle \langle \zeta_{\mathbf{p-q}} \zeta_{\mathbf{k}'-\mathbf{p}'} \rangle$]{\includegraphics[width=.33\columnwidth]{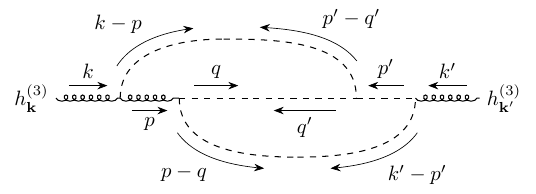}}
        \subfloat[$\langle \zeta_{\mathbf{k}-\mathbf{p}}\zeta_{\mathbf{p}'-\mathbf{q}'} \rangle  \langle \zeta_{\mathbf{p-q}}\zeta_{\mathbf{q}'} \rangle \langle \zeta_{\mathbf{q}} \zeta_{\mathbf{k}'-\mathbf{p}'}\rangle $]{\includegraphics[width=.33\columnwidth]{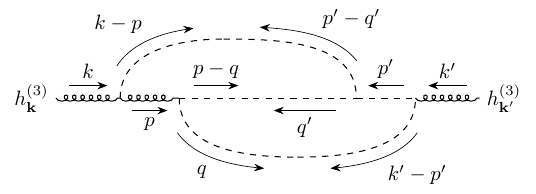}} 
        
\caption{\label{fig:FeynDiag33-2} The two-point function $\langle  h^{\lambda,(3)}_{\mathbf{k},\phi h_{\phi\phi}} h^{\lambda',(3)}_{\mathbf{k}',\phi\psi_{\phi\phi}}\rangle$.}
\end{figure*}

\begin{figure*}[htbp]
    \captionsetup{
      justification=raggedright,
      singlelinecheck=true
    }
    \centering

	\subfloat[$ \langle \zeta_{\mathbf{p-q}}\zeta_{\mathbf{p}'-\mathbf{q}'} \rangle \langle \zeta_{\mathbf{q}}\zeta_{\mathbf{q}'} \rangle \langle \zeta_{\mathbf{k}-\mathbf{p}}\zeta_{\mathbf{k}'-\mathbf{p}'} \rangle $]{\includegraphics[width=.33\columnwidth]{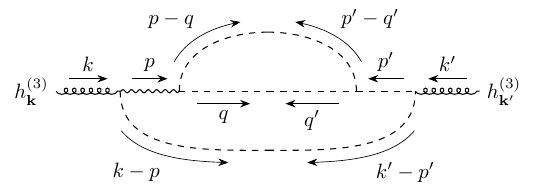}}
        \subfloat[$\langle \zeta_{\mathbf{k}-\mathbf{p}}\zeta_{\mathbf{q}'} \rangle  \langle \zeta_{\mathbf{p-q}}\zeta_{\mathbf{p}'-\mathbf{q}'} \rangle \langle \zeta_{\mathbf{q}}\zeta_{\mathbf{k}'-\mathbf{p}'} \rangle $]{\includegraphics[width=.33\columnwidth]{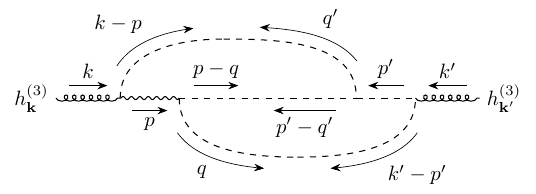}}
        \subfloat[$\langle \zeta_{\mathbf{k}-\mathbf{p}}\zeta_{\mathbf{q}'} \rangle \langle \zeta_{\mathbf{q}} \zeta_{\mathbf{p}'-\mathbf{q}'}\rangle \langle \zeta_{\mathbf{p-q}}\zeta_{\mathbf{k}'-\mathbf{p}'} \rangle$]{\includegraphics[width=.33\columnwidth]{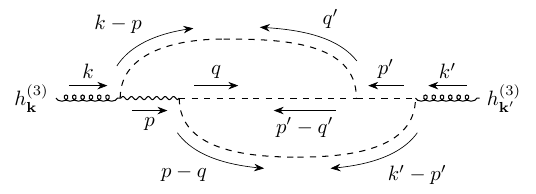}} \\

        \subfloat[$ \langle \zeta_{\mathbf{p-q}}\zeta_{\mathbf{q}'} \rangle \langle \zeta_{\mathbf{q}}\zeta_{\mathbf{p}'-\mathbf{q}'} \rangle \langle \zeta_{\mathbf{k}-\mathbf{p}}\zeta_{\mathbf{k}'-\mathbf{p}'} \rangle $]{\includegraphics[width=.33\columnwidth]{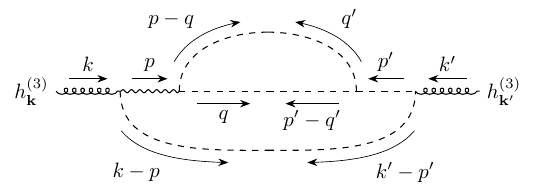}}
        \subfloat[$\langle \zeta_{\mathbf{k}-\mathbf{p}}\zeta_{\mathbf{p}'-\mathbf{q}'} \rangle \langle \zeta_{\mathbf{q}} \zeta_{\mathbf{q}'}\rangle \langle \zeta_{\mathbf{p-q}} \zeta_{\mathbf{k}'-\mathbf{p}'} \rangle$]{\includegraphics[width=.33\columnwidth]{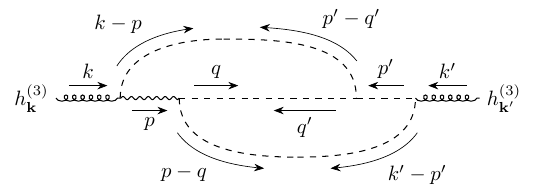}}
        \subfloat[$\langle \zeta_{\mathbf{k}-\mathbf{p}}\zeta_{\mathbf{p}'-\mathbf{q}'} \rangle  \langle \zeta_{\mathbf{p-q}}\zeta_{\mathbf{q}'} \rangle \langle \zeta_{\mathbf{q}} \zeta_{\mathbf{k}'-\mathbf{p}'}\rangle $]{\includegraphics[width=.33\columnwidth]{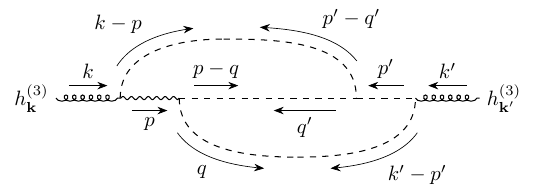}} 
        
\caption{\label{fig:FeynDiag33-3} The two-point function $\langle  h^{\lambda,(3)}_{\mathbf{k},\phi V_{\phi\phi}} h^{\lambda',(3)}_{\mathbf{k}',\phi\psi_{\phi\phi}}\rangle$.}
\end{figure*}

\begin{figure*}[htbp]
    \captionsetup{
      justification=raggedright,
      singlelinecheck=true
    }
    \centering

	\subfloat[$ \langle \zeta_{\mathbf{p-q}}\zeta_{\mathbf{p}'-\mathbf{q}'} \rangle \langle \zeta_{\mathbf{q}}\zeta_{\mathbf{q}'} \rangle \langle \zeta_{\mathbf{k}-\mathbf{p}}\zeta_{\mathbf{k}'-\mathbf{p}'} \rangle $]{\includegraphics[width=.33\columnwidth]{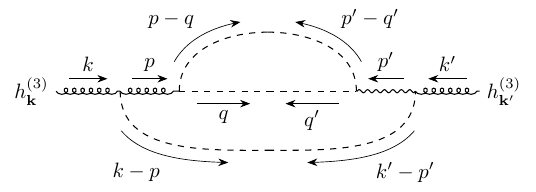}}
        \subfloat[$\langle \zeta_{\mathbf{k}-\mathbf{p}}\zeta_{\mathbf{q}'} \rangle  \langle \zeta_{\mathbf{p-q}}\zeta_{\mathbf{p}'-\mathbf{q}'} \rangle \langle \zeta_{\mathbf{q}}\zeta_{\mathbf{k}'-\mathbf{p}'} \rangle $]{\includegraphics[width=.33\columnwidth]{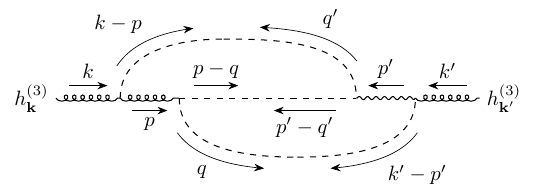}}
        \subfloat[$\langle \zeta_{\mathbf{k}-\mathbf{p}}\zeta_{\mathbf{q}'} \rangle \langle \zeta_{\mathbf{q}} \zeta_{\mathbf{p}'-\mathbf{q}'}\rangle \langle \zeta_{\mathbf{p-q}}\zeta_{\mathbf{k}'-\mathbf{p}'} \rangle$]{\includegraphics[width=.33\columnwidth]{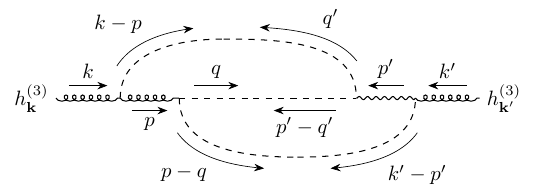}} \\

        \subfloat[$ \langle \zeta_{\mathbf{p-q}}\zeta_{\mathbf{q}'} \rangle \langle \zeta_{\mathbf{q}}\zeta_{\mathbf{p}'-\mathbf{q}'} \rangle \langle \zeta_{\mathbf{k}-\mathbf{p}}\zeta_{\mathbf{k}'-\mathbf{p}'} \rangle $]{\includegraphics[width=.33\columnwidth]{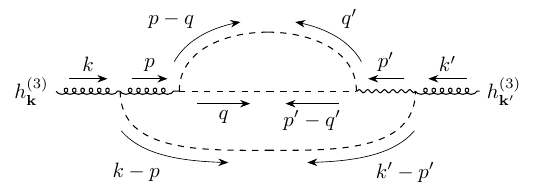}}
        \subfloat[$\langle \zeta_{\mathbf{k}-\mathbf{p}}\zeta_{\mathbf{p}'-\mathbf{q}'} \rangle \langle \zeta_{\mathbf{q}} \zeta_{\mathbf{q}'}\rangle \langle \zeta_{\mathbf{p-q}} \zeta_{\mathbf{k}'-\mathbf{p}'} \rangle$]{\includegraphics[width=.33\columnwidth]{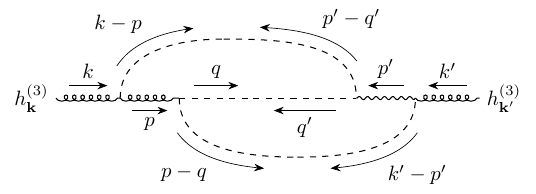}}
        \subfloat[$\langle \zeta_{\mathbf{k}-\mathbf{p}}\zeta_{\mathbf{p}'-\mathbf{q}'} \rangle  \langle \zeta_{\mathbf{p-q}}\zeta_{\mathbf{q}'} \rangle \langle \zeta_{\mathbf{q}} \zeta_{\mathbf{k}'-\mathbf{p}'}\rangle $]{\includegraphics[width=.33\columnwidth]{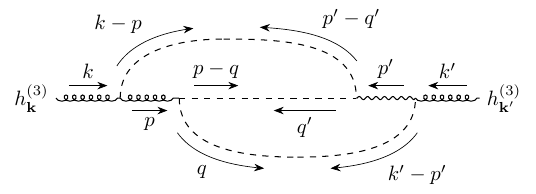}} 
        
\caption{\label{fig:FeynDiag33-4} The two-point function $\langle  h^{\lambda,(3)}_{\mathbf{k},\phi h_{\phi\phi}} h^{\lambda',(3)}_{\mathbf{k}',\phi V_{\phi\phi}}\rangle$.}
\end{figure*}

\begin{figure*}[htbp]
    \captionsetup{
      justification=raggedright,
      singlelinecheck=true
    }
    \centering

	\subfloat[$ \langle \zeta_{\mathbf{p-q}}\zeta_{\mathbf{p}'-\mathbf{q}'} \rangle \langle \zeta_{\mathbf{q}}\zeta_{\mathbf{q}'} \rangle \langle \zeta_{\mathbf{k}-\mathbf{p}}\zeta_{\mathbf{k}'-\mathbf{p}'} \rangle $]{\includegraphics[width=.33\columnwidth]{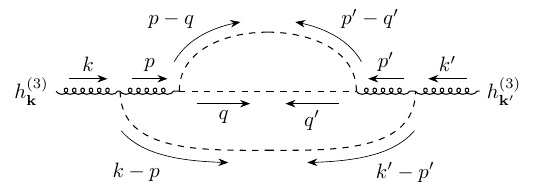}}
        \subfloat[$\langle \zeta_{\mathbf{k}-\mathbf{p}}\zeta_{\mathbf{q}'} \rangle  \langle \zeta_{\mathbf{p-q}}\zeta_{\mathbf{p}'-\mathbf{q}'} \rangle \langle \zeta_{\mathbf{q}}\zeta_{\mathbf{k}'-\mathbf{p}'} \rangle $]{\includegraphics[width=.33\columnwidth]{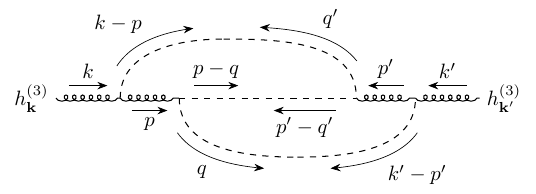}}
        \subfloat[$\langle \zeta_{\mathbf{k}-\mathbf{p}}\zeta_{\mathbf{q}'} \rangle \langle \zeta_{\mathbf{q}} \zeta_{\mathbf{p}'-\mathbf{q}'}\rangle \langle \zeta_{\mathbf{p-q}}\zeta_{\mathbf{k}'-\mathbf{p}'} \rangle$]{\includegraphics[width=.33\columnwidth]{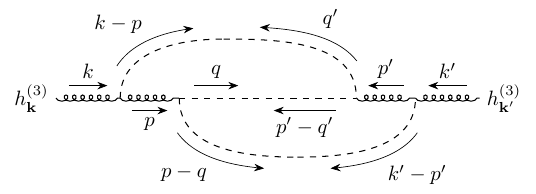}} \\

        \subfloat[$ \langle \zeta_{\mathbf{p-q}}\zeta_{\mathbf{q}'} \rangle \langle \zeta_{\mathbf{q}}\zeta_{\mathbf{p}'-\mathbf{q}'} \rangle \langle \zeta_{\mathbf{k}-\mathbf{p}}\zeta_{\mathbf{k}'-\mathbf{p}'} \rangle $]{\includegraphics[width=.33\columnwidth]{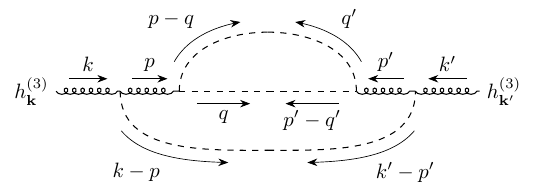}}
        \subfloat[$\langle \zeta_{\mathbf{k}-\mathbf{p}}\zeta_{\mathbf{p}'-\mathbf{q}'} \rangle \langle \zeta_{\mathbf{q}} \zeta_{\mathbf{q}'}\rangle \langle \zeta_{\mathbf{p-q}} \zeta_{\mathbf{k}'-\mathbf{p}'} \rangle$]{\includegraphics[width=.33\columnwidth]{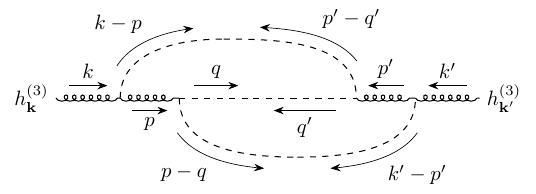}}
        \subfloat[$\langle \zeta_{\mathbf{k}-\mathbf{p}}\zeta_{\mathbf{p}'-\mathbf{q}'} \rangle  \langle \zeta_{\mathbf{p-q}}\zeta_{\mathbf{q}'} \rangle \langle \zeta_{\mathbf{q}} \zeta_{\mathbf{k}'-\mathbf{p}'}\rangle $]{\includegraphics[width=.33\columnwidth]{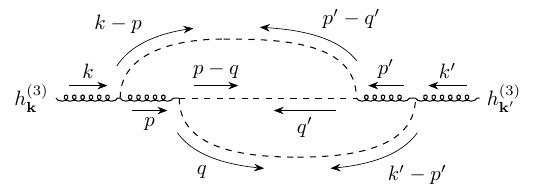}} 
        
\caption{\label{fig:FeynDiag33-5} The two-point function $\langle  h^{\lambda,(3)}_{\mathbf{k},\phi h_{\phi\phi}} h^{\lambda',(3)}_{\mathbf{k}',\phi h_{\phi\phi}}\rangle$.}
\end{figure*}

\begin{figure*}[htbp]
    \captionsetup{
      justification=raggedright,
      singlelinecheck=true
    }
    \centering

	\subfloat[$ \langle \zeta_{\mathbf{p-q}}\zeta_{\mathbf{p}'-\mathbf{q}'} \rangle \langle \zeta_{\mathbf{q}}\zeta_{\mathbf{q}'} \rangle \langle \zeta_{\mathbf{k}-\mathbf{p}}\zeta_{\mathbf{k}'-\mathbf{p}'} \rangle $]{\includegraphics[width=.33\columnwidth]{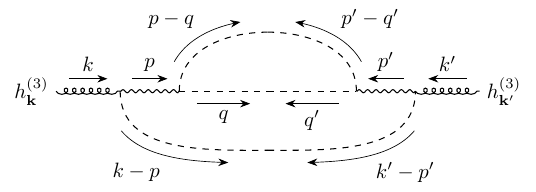}}
        \subfloat[$\langle \zeta_{\mathbf{k}-\mathbf{p}}\zeta_{\mathbf{q}'} \rangle  \langle \zeta_{\mathbf{p-q}}\zeta_{\mathbf{p}'-\mathbf{q}'} \rangle \langle \zeta_{\mathbf{q}}\zeta_{\mathbf{k}'-\mathbf{p}'} \rangle $]{\includegraphics[width=.33\columnwidth]{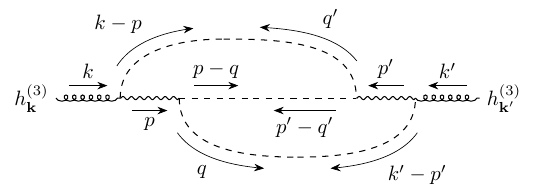}}
        \subfloat[$\langle \zeta_{\mathbf{k}-\mathbf{p}}\zeta_{\mathbf{q}'} \rangle \langle \zeta_{\mathbf{q}} \zeta_{\mathbf{p}'-\mathbf{q}'}\rangle \langle \zeta_{\mathbf{p-q}}\zeta_{\mathbf{k}'-\mathbf{p}'} \rangle$]{\includegraphics[width=.33\columnwidth]{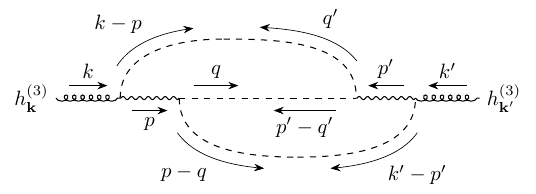}} \\

        \subfloat[$ \langle \zeta_{\mathbf{p-q}}\zeta_{\mathbf{q}'} \rangle \langle \zeta_{\mathbf{q}}\zeta_{\mathbf{p}'-\mathbf{q}'} \rangle \langle \zeta_{\mathbf{k}-\mathbf{p}}\zeta_{\mathbf{k}'-\mathbf{p}'} \rangle $]{\includegraphics[width=.33\columnwidth]{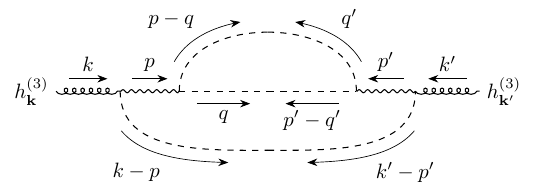}}
        \subfloat[$\langle \zeta_{\mathbf{k}-\mathbf{p}}\zeta_{\mathbf{p}'-\mathbf{q}'} \rangle \langle \zeta_{\mathbf{q}} \zeta_{\mathbf{q}'}\rangle \langle \zeta_{\mathbf{p-q}} \zeta_{\mathbf{k}'-\mathbf{p}'} \rangle$]{\includegraphics[width=.33\columnwidth]{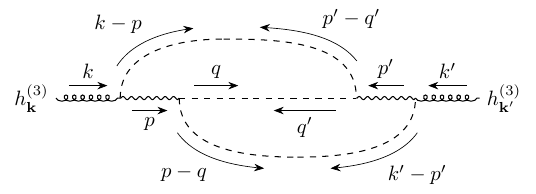}}
        \subfloat[$\langle \zeta_{\mathbf{k}-\mathbf{p}}\zeta_{\mathbf{p}'-\mathbf{q}'} \rangle  \langle \zeta_{\mathbf{p-q}}\zeta_{\mathbf{q}'} \rangle \langle \zeta_{\mathbf{q}} \zeta_{\mathbf{k}'-\mathbf{p}'}\rangle $]{\includegraphics[width=.33\columnwidth]{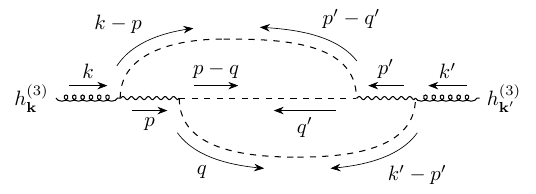}} 
        
\caption{\label{fig:FeynDiag33-6} The two-point function $\langle  h^{\lambda,(3)}_{\mathbf{k},\phi V_{\phi\phi}} h^{\lambda',(3)}_{\mathbf{k}',\phi V_{\phi\phi}}\rangle$.}
\end{figure*}

\begin{figure*}[htbp]
    \captionsetup{
      justification=raggedright,
      singlelinecheck=true
    }
    \centering

	\subfloat[$ \langle \zeta_{\mathbf{p-q}}\zeta_{\mathbf{p}'-\mathbf{q}'} \rangle \langle \zeta_{\mathbf{q}}\zeta_{\mathbf{q}'} \rangle \langle \zeta_{\mathbf{k}-\mathbf{p}}\zeta_{\mathbf{k}'-\mathbf{p}'} \rangle $]{\includegraphics[width=.33\columnwidth]{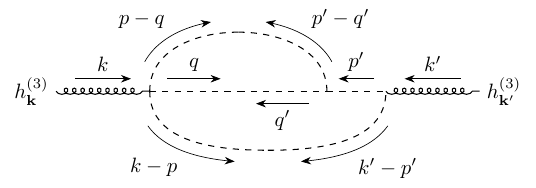}}
        \subfloat[$\langle \zeta_{\mathbf{k}-\mathbf{p}}\zeta_{\mathbf{q}'} \rangle  \langle \zeta_{\mathbf{p-q}}\zeta_{\mathbf{p}'-\mathbf{q}'} \rangle \langle \zeta_{\mathbf{q}}\zeta_{\mathbf{k}'-\mathbf{p}'} \rangle $]{\includegraphics[width=.33\columnwidth]{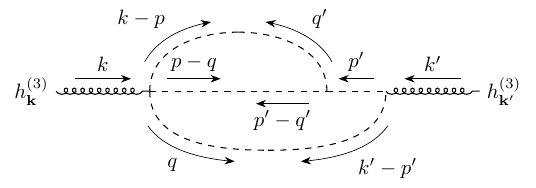}}
        \subfloat[$\langle \zeta_{\mathbf{k}-\mathbf{p}}\zeta_{\mathbf{q}'} \rangle \langle \zeta_{\mathbf{q}} \zeta_{\mathbf{p}'-\mathbf{q}'}\rangle \langle \zeta_{\mathbf{p-q}}\zeta_{\mathbf{k}'-\mathbf{p}'} \rangle$]{\includegraphics[width=.33\columnwidth]{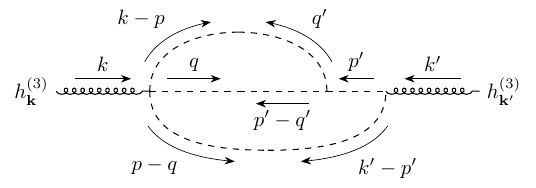}} \\

        \subfloat[$ \langle \zeta_{\mathbf{p-q}}\zeta_{\mathbf{q}'} \rangle \langle \zeta_{\mathbf{q}}\zeta_{\mathbf{p}'-\mathbf{q}'} \rangle \langle \zeta_{\mathbf{k}-\mathbf{p}}\zeta_{\mathbf{k}'-\mathbf{p}'} \rangle $]{\includegraphics[width=.33\columnwidth]{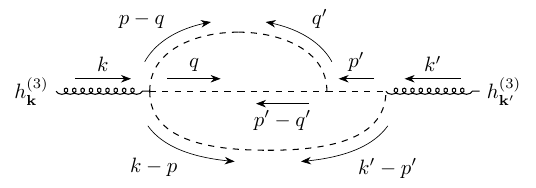}}
        \subfloat[$\langle \zeta_{\mathbf{k}-\mathbf{p}}\zeta_{\mathbf{p}'-\mathbf{q}'} \rangle \langle \zeta_{\mathbf{q}} \zeta_{\mathbf{q}'}\rangle \langle \zeta_{\mathbf{p-q}} \zeta_{\mathbf{k}'-\mathbf{p}'} \rangle$]{\includegraphics[width=.33\columnwidth]{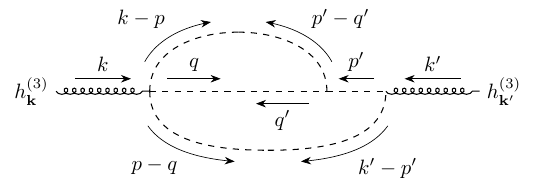}}
        \subfloat[$\langle \zeta_{\mathbf{k}-\mathbf{p}}\zeta_{\mathbf{p}'-\mathbf{q}'} \rangle  \langle \zeta_{\mathbf{p-q}}\zeta_{\mathbf{q}'} \rangle \langle \zeta_{\mathbf{q}} \zeta_{\mathbf{k}'-\mathbf{p}'}\rangle $]{\includegraphics[width=.33\columnwidth]{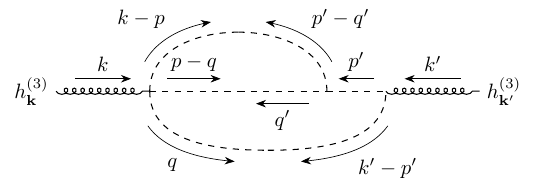}} 
        
\caption{\label{fig:FeynDiag33-7} The two-point function $\langle  h^{\lambda,(3)}_{\mathbf{k},\phi\phi\phi} h^{\lambda',(3)}_{\mathbf{k}',\phi \psi_{\phi\phi}}\rangle$.}
\end{figure*}

\begin{figure*}[htbp]
    \captionsetup{
      justification=raggedright,
      singlelinecheck=true
    }
    \centering

	\subfloat[$ \langle \zeta_{\mathbf{p-q}}\zeta_{\mathbf{p}'-\mathbf{q}'} \rangle \langle \zeta_{\mathbf{q}}\zeta_{\mathbf{q}'} \rangle \langle \zeta_{\mathbf{k}-\mathbf{p}}\zeta_{\mathbf{k}'-\mathbf{p}'} \rangle $]{\includegraphics[width=.33\columnwidth]{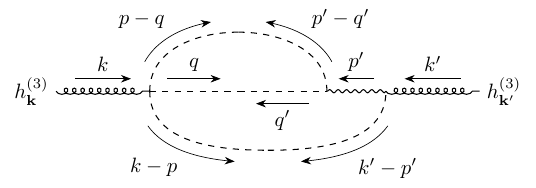}}
        \subfloat[$\langle \zeta_{\mathbf{k}-\mathbf{p}}\zeta_{\mathbf{q}'} \rangle  \langle \zeta_{\mathbf{p-q}}\zeta_{\mathbf{p}'-\mathbf{q}'} \rangle \langle \zeta_{\mathbf{q}}\zeta_{\mathbf{k}'-\mathbf{p}'} \rangle $]{\includegraphics[width=.33\columnwidth]{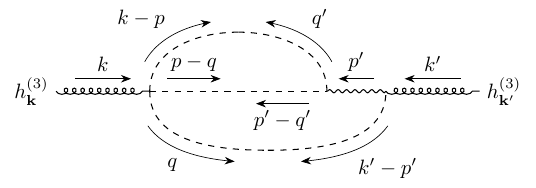}}
        \subfloat[$\langle \zeta_{\mathbf{k}-\mathbf{p}}\zeta_{\mathbf{q}'} \rangle \langle \zeta_{\mathbf{q}} \zeta_{\mathbf{p}'-\mathbf{q}'}\rangle \langle \zeta_{\mathbf{p-q}}\zeta_{\mathbf{k}'-\mathbf{p}'} \rangle$]{\includegraphics[width=.33\columnwidth]{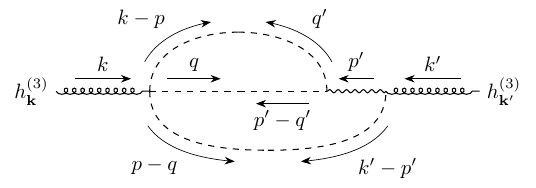}} \\

        \subfloat[$ \langle \zeta_{\mathbf{p-q}}\zeta_{\mathbf{q}'} \rangle \langle \zeta_{\mathbf{q}}\zeta_{\mathbf{p}'-\mathbf{q}'} \rangle \langle \zeta_{\mathbf{k}-\mathbf{p}}\zeta_{\mathbf{k}'-\mathbf{p}'} \rangle $]{\includegraphics[width=.33\columnwidth]{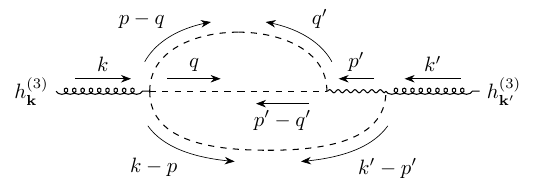}}
        \subfloat[$\langle \zeta_{\mathbf{k}-\mathbf{p}}\zeta_{\mathbf{p}'-\mathbf{q}'} \rangle \langle \zeta_{\mathbf{q}} \zeta_{\mathbf{q}'}\rangle \langle \zeta_{\mathbf{p-q}} \zeta_{\mathbf{k}'-\mathbf{p}'} \rangle$]{\includegraphics[width=.33\columnwidth]{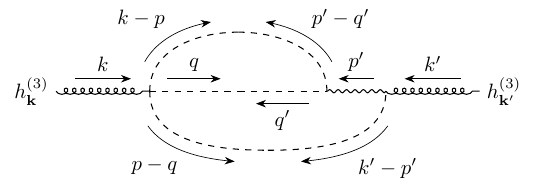}}
        \subfloat[$\langle \zeta_{\mathbf{k}-\mathbf{p}}\zeta_{\mathbf{p}'-\mathbf{q}'} \rangle  \langle \zeta_{\mathbf{p-q}}\zeta_{\mathbf{q}'} \rangle \langle \zeta_{\mathbf{q}} \zeta_{\mathbf{k}'-\mathbf{p}'}\rangle $]{\includegraphics[width=.33\columnwidth]{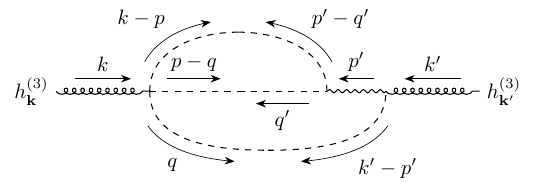}} 
        
\caption{\label{fig:FeynDiag33-8} The two-point function $\langle  h^{\lambda,(3)}_{\mathbf{k},\phi\phi\phi} h^{\lambda',(3)}_{\mathbf{k}',\phi V_{\phi\phi}}\rangle$.}
\end{figure*}

\begin{figure*}[htbp]
    \captionsetup{
      justification=raggedright,
      singlelinecheck=true
    }
    \centering

	\subfloat[$ \langle \zeta_{\mathbf{p-q}}\zeta_{\mathbf{p}'-\mathbf{q}'} \rangle \langle \zeta_{\mathbf{q}}\zeta_{\mathbf{q}'} \rangle \langle \zeta_{\mathbf{k}-\mathbf{p}}\zeta_{\mathbf{k}'-\mathbf{p}'} \rangle $]{\includegraphics[width=.33\columnwidth]{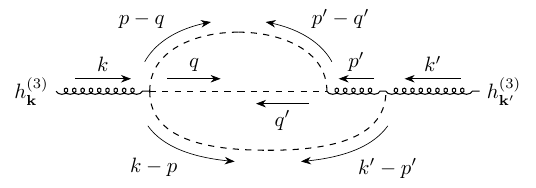}}
        \subfloat[$\langle \zeta_{\mathbf{k}-\mathbf{p}}\zeta_{\mathbf{q}'} \rangle  \langle \zeta_{\mathbf{p-q}}\zeta_{\mathbf{p}'-\mathbf{q}'} \rangle \langle \zeta_{\mathbf{q}}\zeta_{\mathbf{k}'-\mathbf{p}'} \rangle $]{\includegraphics[width=.33\columnwidth]{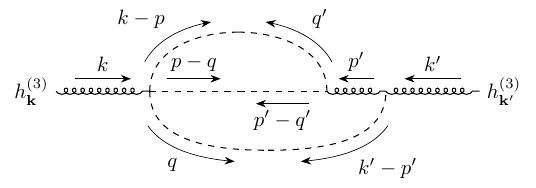}}
        \subfloat[$\langle \zeta_{\mathbf{k}-\mathbf{p}}\zeta_{\mathbf{q}'} \rangle \langle \zeta_{\mathbf{q}} \zeta_{\mathbf{p}'-\mathbf{q}'}\rangle \langle \zeta_{\mathbf{p-q}}\zeta_{\mathbf{k}'-\mathbf{p}'} \rangle$]{\includegraphics[width=.33\columnwidth]{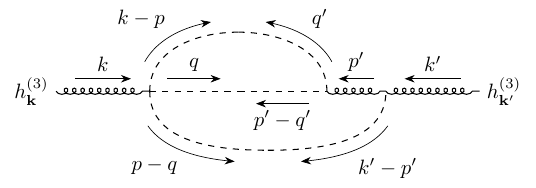}} \\

        \subfloat[$ \langle \zeta_{\mathbf{p-q}}\zeta_{\mathbf{q}'} \rangle \langle \zeta_{\mathbf{q}}\zeta_{\mathbf{p}'-\mathbf{q}'} \rangle \langle \zeta_{\mathbf{k}-\mathbf{p}}\zeta_{\mathbf{k}'-\mathbf{p}'} \rangle $]{\includegraphics[width=.33\columnwidth]{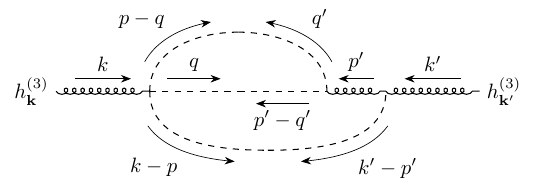}}
        \subfloat[$\langle \zeta_{\mathbf{k}-\mathbf{p}}\zeta_{\mathbf{p}'-\mathbf{q}'} \rangle \langle \zeta_{\mathbf{q}} \zeta_{\mathbf{q}'}\rangle \langle \zeta_{\mathbf{p-q}} \zeta_{\mathbf{k}'-\mathbf{p}'} \rangle$]{\includegraphics[width=.33\columnwidth]{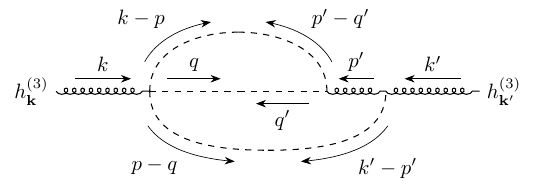}}
        \subfloat[$\langle \zeta_{\mathbf{k}-\mathbf{p}}\zeta_{\mathbf{p}'-\mathbf{q}'} \rangle  \langle \zeta_{\mathbf{p-q}}\zeta_{\mathbf{q}'} \rangle \langle \zeta_{\mathbf{q}} \zeta_{\mathbf{k}'-\mathbf{p}'}\rangle $]{\includegraphics[width=.33\columnwidth]{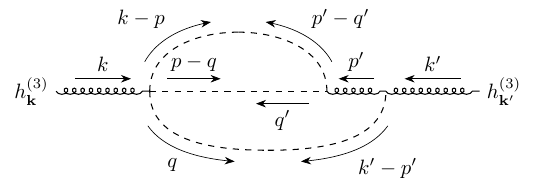}} 
        
\caption{\label{fig:FeynDiag33-9}The two-point function $\langle  h^{\lambda,(3)}_{\mathbf{k},\phi\phi\phi} h^{\lambda',(3)}_{\mathbf{k}',\phi h_{\phi\phi}}\rangle$.}
\end{figure*}

\begin{figure*}[htbp]
    \captionsetup{
      justification=raggedright,
      singlelinecheck=true
    }
    \centering

	\subfloat[$ \langle \zeta_{\mathbf{p-q}}\zeta_{\mathbf{p}'-\mathbf{q}'} \rangle \langle \zeta_{\mathbf{q}}\zeta_{\mathbf{q}'} \rangle \langle \zeta_{\mathbf{k}-\mathbf{p}}\zeta_{\mathbf{k}'-\mathbf{p}'} \rangle $]{\includegraphics[width=.33\columnwidth]{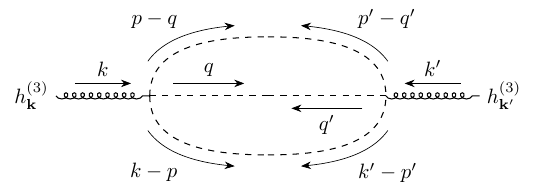}}
        \subfloat[$\langle \zeta_{\mathbf{k}-\mathbf{p}}\zeta_{\mathbf{q}'} \rangle  \langle \zeta_{\mathbf{p-q}}\zeta_{\mathbf{p}'-\mathbf{q}'} \rangle \langle \zeta_{\mathbf{q}}\zeta_{\mathbf{k}'-\mathbf{p}'} \rangle $]{\includegraphics[width=.33\columnwidth]{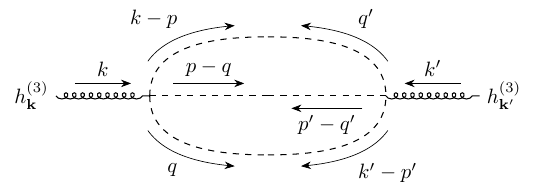}}
        \subfloat[$\langle \zeta_{\mathbf{k}-\mathbf{p}}\zeta_{\mathbf{q}'} \rangle \langle \zeta_{\mathbf{q}} \zeta_{\mathbf{p}'-\mathbf{q}'}\rangle \langle \zeta_{\mathbf{p-q}}\zeta_{\mathbf{k}'-\mathbf{p}'} \rangle$]{\includegraphics[width=.33\columnwidth]{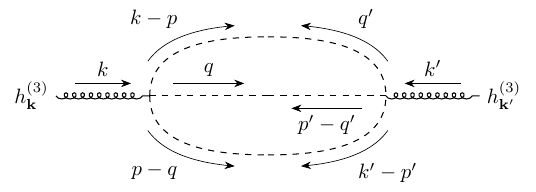}} \\

        \subfloat[$ \langle \zeta_{\mathbf{p-q}}\zeta_{\mathbf{q}'} \rangle \langle \zeta_{\mathbf{q}}\zeta_{\mathbf{p}'-\mathbf{q}'} \rangle \langle \zeta_{\mathbf{k}-\mathbf{p}}\zeta_{\mathbf{k}'-\mathbf{p}'} \rangle $]{\includegraphics[width=.33\columnwidth]{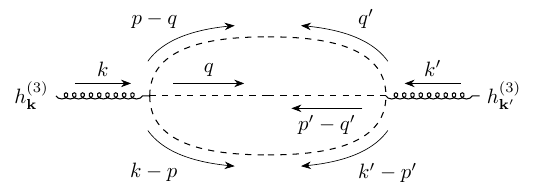}}
        \subfloat[$\langle \zeta_{\mathbf{k}-\mathbf{p}}\zeta_{\mathbf{p}'-\mathbf{q}'} \rangle \langle \zeta_{\mathbf{q}} \zeta_{\mathbf{q}'}\rangle \langle \zeta_{\mathbf{p-q}} \zeta_{\mathbf{k}'-\mathbf{p}'} \rangle$]{\includegraphics[width=.33\columnwidth]{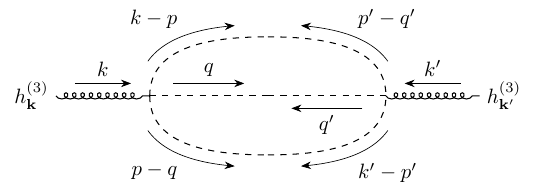}}
        \subfloat[$\langle \zeta_{\mathbf{k}-\mathbf{p}}\zeta_{\mathbf{p}'-\mathbf{q}'} \rangle  \langle \zeta_{\mathbf{p-q}}\zeta_{\mathbf{q}'} \rangle \langle \zeta_{\mathbf{q}} \zeta_{\mathbf{k}'-\mathbf{p}'}\rangle $]{\includegraphics[width=.33\columnwidth]{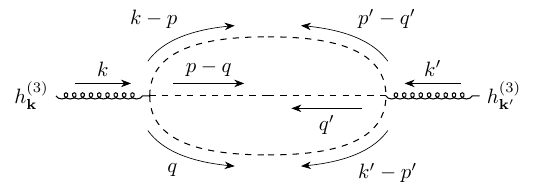}} 
        
\caption{\label{fig:FeynDiag33-10} The two-point function $\langle  h^{\lambda,(3)}_{\mathbf{k},\phi\phi\phi} h^{\lambda',(3)}_{\mathbf{k}',\phi \phi\phi}\rangle$.}
\end{figure*}

\

\

\

\

\

\bibliography{biblio}

\providecommand{\href}[2]{#2}\begingroup\raggedright\begin{thebibliography}{10}

\bibitem{Hindmarsh:2017gnf}
M.~Hindmarsh, S.J.~Huber, K.~Rummukainen and D.J.~Weir, \emph{{Shape of the
  acoustic gravitational wave power spectrum from a first order phase
  transition}}, \href{https://doi.org/10.1103/PhysRevD.96.103520}{\emph{Phys.
  Rev. D} {\bfseries 96} (2017) 103520}
  [\href{https://arxiv.org/abs/1704.05871}{{\ttfamily 1704.05871}}].

\bibitem{Guo:2020grp}
H.-K.~Guo, K.~Sinha, D.~Vagie and G.~White, \emph{{Phase Transitions in an
  Expanding Universe: Stochastic Gravitational Waves in Standard and
  Non-Standard Histories}},
  \href{https://doi.org/10.1088/1475-7516/2021/01/001}{\emph{JCAP} {\bfseries
  01} (2021) 001} [\href{https://arxiv.org/abs/2007.08537}{{\ttfamily
  2007.08537}}].

\bibitem{Hiramatsu:2010yz}
T.~Hiramatsu, M.~Kawasaki and K.~Saikawa, \emph{{Gravitational Waves from
  Collapsing Domain Walls}},
  \href{https://doi.org/10.1088/1475-7516/2010/05/032}{\emph{JCAP} {\bfseries
  05} (2010) 032} [\href{https://arxiv.org/abs/1002.1555}{{\ttfamily
  1002.1555}}].

\bibitem{Blanco-Pillado:2011egf}
J.J.~Blanco-Pillado, K.D.~Olum and B.~Shlaer, \emph{{Large parallel cosmic
  string simulations: New results on loop production}},
  \href{https://doi.org/10.1103/PhysRevD.83.083514}{\emph{Phys. Rev. D}
  {\bfseries 83} (2011) 083514}
  [\href{https://arxiv.org/abs/1101.5173}{{\ttfamily 1101.5173}}].

\bibitem{Kawasaki:2011vv}
M.~Kawasaki and K.~Saikawa, \emph{{Study of gravitational radiation from cosmic
  domain walls}},
  \href{https://doi.org/10.1088/1475-7516/2011/09/008}{\emph{JCAP} {\bfseries
  09} (2011) 008} [\href{https://arxiv.org/abs/1102.5628}{{\ttfamily
  1102.5628}}].

\bibitem{Rajagopal:1994zj}
M.~Rajagopal and R.W.~Romani, \emph{{Ultralow frequency gravitational radiation
  from massive black hole binaries}},
  \href{https://doi.org/10.1086/175813}{\emph{Astrophys. J.} {\bfseries 446}
  (1995) 543} [\href{https://arxiv.org/abs/astro-ph/9412038}{{\ttfamily
  astro-ph/9412038}}].

\bibitem{Kocsis:2010xa}
B.~Kocsis and A.~Sesana, \emph{{Gas driven massive black hole binaries:
  signatures in the nHz gravitational wave background}},
  \href{https://doi.org/10.1111/j.1365-2966.2010.17782.x}{\emph{Mon. Not. Roy.
  Astron. Soc.} {\bfseries 411} (2011) 1467}
  [\href{https://arxiv.org/abs/1002.0584}{{\ttfamily 1002.0584}}].

\bibitem{Ananda_2007}
K.N.~Ananda, C.~Clarkson and D.~Wands, \emph{Cosmological gravitational wave
  background from primordial density perturbations},
  \href{https://doi.org/10.1103/physrevd.75.123518}{\emph{Physical Review D}
  {\bfseries 75} (2007) }.

\bibitem{Kohri:2018awv}
K.~Kohri and T.~Terada, \emph{{Semianalytic calculation of gravitational wave
  spectrum nonlinearly induced from primordial curvature perturbations}},
  \href{https://doi.org/10.1103/PhysRevD.97.123532}{\emph{Phys. Rev. D}
  {\bfseries 97} (2018) 123532}
  [\href{https://arxiv.org/abs/1804.08577}{{\ttfamily 1804.08577}}].

\bibitem{NANOGrav:2023gor}
{\scshape NANOGrav} collaboration, \emph{{The NANOGrav 15 yr Data Set: Evidence
  for a Gravitational-wave Background}},
  \href{https://doi.org/10.3847/2041-8213/acdac6}{\emph{Astrophys. J. Lett.}
  {\bfseries 951} (2023) L8}
  [\href{https://arxiv.org/abs/2306.16213}{{\ttfamily 2306.16213}}].

\bibitem{NANOGrav:2023hde}
{\scshape NANOGrav} collaboration, \emph{{The NANOGrav 15 yr Data Set:
  Observations and Timing of 68 Millisecond Pulsars}},
  \href{https://doi.org/10.3847/2041-8213/acda9a}{\emph{Astrophys. J. Lett.}
  {\bfseries 951} (2023) L9}
  [\href{https://arxiv.org/abs/2306.16217}{{\ttfamily 2306.16217}}].

\bibitem{NANOGrav:2023hvm}
{\scshape NANOGrav} collaboration, \emph{{The NANOGrav 15 yr Data Set: Search
  for Signals from New Physics}},
  \href{https://doi.org/10.3847/2041-8213/acdc91}{\emph{Astrophys. J. Lett.}
  {\bfseries 951} (2023) L11}
  [\href{https://arxiv.org/abs/2306.16219}{{\ttfamily 2306.16219}}].

\bibitem{EPTA:2023_1}
{\scshape EPTA} collaboration, \emph{{The second data release from the European
  Pulsar Timing Array I. The dataset and timing analysis}},
  \href{https://arxiv.org/abs/2306.16224}{{\ttfamily 2306.16224}}.

\bibitem{EPTA:2023_2}
J.~Antoniadis, P.~Arumugam, S.~Arumugam, S.~Babak, M.~Bagchi, A.S.B.~Nielsen
  et~al., \emph{The second data release from the european pulsar timing array
  ii. customised pulsar noise models for spatially correlated gravitational
  waves},  2023.

\bibitem{EPTA:2023_3}
J.~Antoniadis, P.~Arumugam, S.~Arumugam, S.~Babak, M.~Bagchi, A.S.B.~Nielsen
  et~al., \emph{The second data release from the european pulsar timing array
  iii. search for gravitational wave signals},  2023.

\bibitem{EPTA:2023_4}
J.~Antoniadis, P.~Arumugam, S.~Arumugam, S.~Babak, M.~Bagchi, A.S.B.~Nielsen
  et~al., \emph{The second data release from the european pulsar timing array
  iv. search for continuous gravitational wave signals},  2023.

\bibitem{EPTA:2023_5}
J.~Antoniadis, P.~Arumugam, S.~Arumugam, P.~Auclair, S.~Babak, M.~Bagchi
  et~al., \emph{The second data release from the european pulsar timing array:
  V. implications for massive black holes, dark matter and the early universe},
   2023.

\bibitem{EPTA:2023_6}
C.~Smarra, B.~Goncharov, E.~Barausse, J.~Antoniadis, S.~Babak, A.S.B.~Nielsen
  et~al., \emph{The second data release from the european pulsar timing array:
  Vi. challenging the ultralight dark matter paradigm},  2023.

\bibitem{PPTA:2023_1}
D.J.~Reardon, A.~Zic, R.M.~Shannon, G.B.~Hobbs, M.~Bailes, V.D.~Marco et~al.,
  \emph{Search for an isotropic gravitational-wave background with the parkes
  pulsar timing array},
  \href{https://doi.org/10.3847/2041-8213/acdd02}{\emph{The Astrophysical
  Journal Letters} {\bfseries 951} (2023) L6}.

\bibitem{PPTA:2023_2}
D.J.~Reardon, A.~Zic, R.M.~Shannon, V.D.~Marco, G.B.~Hobbs, A.~Kapur et~al.,
  \emph{The gravitational-wave background null hypothesis: Characterizing noise
  in millisecond pulsar arrival times with the parkes pulsar timing array},
  \href{https://doi.org/10.3847/2041-8213/acdd03}{\emph{The Astrophysical
  Journal Letters} {\bfseries 951} (2023) L7}.

\bibitem{PPTA:2023_3}
A.~Zic, D.J.~Reardon, A.~Kapur, G.~Hobbs, R.~Mandow, M.~Curyło et~al.,
  \emph{The parkes pulsar timing array third data release},  2023.

\bibitem{CPTA:2023}
H.~Xu et~al., \emph{{Searching for the Nano-Hertz Stochastic Gravitational Wave
  Background with the Chinese Pulsar Timing Array Data Release I}},
  \href{https://doi.org/10.1088/1674-4527/acdfa5}{\emph{Res. Astron.
  Astrophys.} {\bfseries 23} (2023) 075024}
  [\href{https://arxiv.org/abs/2306.16216}{{\ttfamily 2306.16216}}].

\bibitem{Ellis:2023tsl}
J.~Ellis, M.~Lewicki, C.~Lin and V.~Vaskonen, \emph{{Cosmic Superstrings
  Revisited in Light of NANOGrav 15-Year Data}},
  \href{https://arxiv.org/abs/2306.17147}{{\ttfamily 2306.17147}}.

\bibitem{Kitajima:2023cek}
N.~Kitajima, J.~Lee, K.~Murai, F.~Takahashi and W.~Yin, \emph{{Nanohertz
  Gravitational Waves from Axion Domain Walls Coupled to QCD}},
  \href{https://arxiv.org/abs/2306.17146}{{\ttfamily 2306.17146}}.

\bibitem{Bai:2023cqj}
Y.~Bai, T.-K.~Chen and M.~Korwar, \emph{{QCD-Collapsed Domain Walls: QCD Phase
  Transition and Gravitational Wave Spectroscopy}},
  \href{https://arxiv.org/abs/2306.17160}{{\ttfamily 2306.17160}}.

\bibitem{Fujikura:2023lkn}
K.~Fujikura, S.~Girmohanta, Y.~Nakai and M.~Suzuki, \emph{{NANOGrav Signal from
  a Dark Conformal Phase Transition}},
  \href{https://arxiv.org/abs/2306.17086}{{\ttfamily 2306.17086}}.

\bibitem{Bringmann:2023opz}
T.~Bringmann, P.F.~Depta, T.~Konstandin, K.~Schmidt-Hoberg and C.~Tasillo,
  \emph{{Does NANOGrav observe a dark sector phase transition?}},
  \href{https://arxiv.org/abs/2306.09411}{{\ttfamily 2306.09411}}.

\bibitem{Depta:2023qst}
P.F.~Depta, K.~Schmidt-Hoberg, P.~Schwaller and C.~Tasillo, \emph{{Do pulsar
  timing arrays observe merging primordial black holes?}},
  \href{https://arxiv.org/abs/2306.17836}{{\ttfamily 2306.17836}}.

\bibitem{Balaji:2023ehk}
S.~Balaji, G.~Dom\`enech and G.~Franciolini, \emph{{Scalar-induced
  gravitational wave interpretation of PTA data: the role of scalar fluctuation
  propagation speed}},  \href{https://arxiv.org/abs/2307.08552}{{\ttfamily
  2307.08552}}.

\bibitem{Cai:2023dls}
Y.-F.~Cai, X.-C.~He, X.~Ma, S.-F.~Yan and G.-W.~Yuan, \emph{{Limits on
  scalar-induced gravitational waves from the stochastic background by pulsar
  timing array observations}},
  \href{https://arxiv.org/abs/2306.17822}{{\ttfamily 2306.17822}}.

\bibitem{wang2023exploring}
S.~Wang, Z.-C.~Zhao, J.-P.~Li and Q.-H.~Zhu, \emph{Exploring the implications
  of 2023 pulsar timing array datasets for scalar-induced gravitational waves
  and primordial black holes},  2023.

\bibitem{Vagnozzi:2023lwo}
S.~Vagnozzi, \emph{{Inflationary interpretation of the stochastic gravitational
  wave background signal detected by pulsar timing array experiments}},
  \href{https://doi.org/10.1016/j.jheap.2023.07.001}{\emph{JHEAp} {\bfseries
  39} (2023) 81} [\href{https://arxiv.org/abs/2306.16912}{{\ttfamily
  2306.16912}}].

\bibitem{Inomata:2023zup}
K.~Inomata, K.~Kohri and T.~Terada, \emph{{The Detected Stochastic
  Gravitational Waves and Subsolar-Mass Primordial Black Holes}},
  \href{https://arxiv.org/abs/2306.17834}{{\ttfamily 2306.17834}}.

\bibitem{Saga:2017hft}
S.~Saga, \emph{{The Vector Mode in the Second-order Cosmological Perturbation
  Theory}}, Ph.D. thesis, Nagoya U. (main), 2017.
\newblock 10.1007/978-981-10-8007-4.

\bibitem{Planck:2018jri}
{\scshape Planck} collaboration, \emph{{Planck 2018 results. X. Constraints on
  inflation}}, \href{https://doi.org/10.1051/0004-6361/201833887}{\emph{Astron.
  Astrophys.} {\bfseries 641} (2020) A10}
  [\href{https://arxiv.org/abs/1807.06211}{{\ttfamily 1807.06211}}].

\bibitem{Adshead:2021hnm}
P.~Adshead, K.D.~Lozanov and Z.J.~Weiner, \emph{{Non-Gaussianity and the
  induced gravitational wave background}},
  \href{https://doi.org/10.1088/1475-7516/2021/10/080}{\emph{JCAP} {\bfseries
  10} (2021) 080} [\href{https://arxiv.org/abs/2105.01659}{{\ttfamily
  2105.01659}}].

\bibitem{Yuan:2023ofl}
C.~Yuan, D.-S.~Meng and Q.-G.~Huang, \emph{{Full analysis of the scalar-induced
  gravitational waves for the curvature perturbation with local-type
  non-Gaussianities}},  \href{https://arxiv.org/abs/2308.07155}{{\ttfamily
  2308.07155}}.

\bibitem{Li:2023xtl}
J.-P.~Li, S.~Wang, Z.-C.~Zhao and K.~Kohri, \emph{{Complete Analysis of
  Scalar-Induced Gravitational Waves and Primordial Non-Gaussianities
  $f_{\mathrm{NL}}$ and $g_{\mathrm{NL}}$}},
  \href{https://arxiv.org/abs/2309.07792}{{\ttfamily 2309.07792}}.

\bibitem{Yuan:2021qgz}
C.~Yuan and Q.-G.~Huang, \emph{{A topic review on probing primordial black hole
  dark matter with scalar induced gravitational waves}},
  \href{https://doi.org/10.1016/j.isci.2021.102860}{\emph{iScience} {\bfseries
  24} (2021) 102860} [\href{https://arxiv.org/abs/2103.04739}{{\ttfamily
  2103.04739}}].

\bibitem{DeLuca:2023tun}
V.~De~Luca, A.~Kehagias and A.~Riotto, \emph{{How well do we know the
  primordial black hole abundance: The crucial role of nonlinearities when
  approaching the horizon}},
  \href{https://doi.org/10.1103/PhysRevD.108.063531}{\emph{Phys. Rev. D}
  {\bfseries 108} (2023) 063531}
  [\href{https://arxiv.org/abs/2307.13633}{{\ttfamily 2307.13633}}].

\bibitem{Hwang:2017oxa}
J.-C.~Hwang, D.~Jeong and H.~Noh, \emph{{Gauge dependence of gravitational
  waves generated from scalar perturbations}},
  \href{https://doi.org/10.3847/1538-4357/aa74be}{\emph{Astrophys. J.}
  {\bfseries 842} (2017) 46}
  [\href{https://arxiv.org/abs/1704.03500}{{\ttfamily 1704.03500}}].

\bibitem{Yuan:2019fwv}
C.~Yuan, Z.-C.~Chen and Q.-G.~Huang, \emph{{Scalar induced gravitational waves
  in different gauges}},
  \href{https://doi.org/10.1103/PhysRevD.101.063018}{\emph{Phys. Rev. D}
  {\bfseries 101} (2020) 063018}
  [\href{https://arxiv.org/abs/1912.00885}{{\ttfamily 1912.00885}}].

\bibitem{Inomata:2019yww}
K.~Inomata and T.~Terada, \emph{{Gauge Independence of Induced Gravitational
  Waves}}, \href{https://doi.org/10.1103/PhysRevD.101.023523}{\emph{Phys. Rev.
  D} {\bfseries 101} (2020) 023523}
  [\href{https://arxiv.org/abs/1912.00785}{{\ttfamily 1912.00785}}].

\bibitem{Domenech:2021and}
G.~Dom\`enech, S.~Passaglia and S.~Renaux-Petel, \emph{{Gravitational waves
  from dark matter isocurvature}},
  \href{https://doi.org/10.1088/1475-7516/2022/03/023}{\emph{JCAP} {\bfseries
  03} (2022) 023} [\href{https://arxiv.org/abs/2112.10163}{{\ttfamily
  2112.10163}}].

\bibitem{Saga:2014jca}
S.~Saga, K.~Ichiki and N.~Sugiyama, \emph{{Impact of anisotropic stress of
  free-streaming particles on gravitational waves induced by cosmological
  density perturbations}},
  \href{https://doi.org/10.1103/PhysRevD.91.024030}{\emph{Phys. Rev. D}
  {\bfseries 91} (2015) 024030}
  [\href{https://arxiv.org/abs/1412.1081}{{\ttfamily 1412.1081}}].

\bibitem{Zhang:2022dgx}
X.~Zhang, J.-Z.~Zhou and Z.~Chang, \emph{{Impact of the free-streaming
  neutrinos to the second order induced gravitational waves}},
  \href{https://doi.org/10.1140/epjc/s10052-022-10742-x}{\emph{Eur. Phys. J. C}
  {\bfseries 82} (2022) 781}
  [\href{https://arxiv.org/abs/2208.12948}{{\ttfamily 2208.12948}}].

\bibitem{Yuan:2019udt}
C.~Yuan, Z.-C.~Chen and Q.-G.~Huang, \emph{{Probing
  primordial\textendash{}black-hole dark matter with scalar induced
  gravitational waves}},
  \href{https://doi.org/10.1103/PhysRevD.100.081301}{\emph{Phys. Rev. D}
  {\bfseries 100} (2019) 081301}
  [\href{https://arxiv.org/abs/1906.11549}{{\ttfamily 1906.11549}}].

\bibitem{Zhou:2021vcw}
J.-Z.~Zhou, X.~Zhang, Q.-H.~Zhu and Z.~Chang, \emph{{The third order scalar
  induced gravitational waves}},
  \href{https://doi.org/10.1088/1475-7516/2022/05/013}{\emph{JCAP} {\bfseries
  05} (2022) 013} [\href{https://arxiv.org/abs/2106.01641}{{\ttfamily
  2106.01641}}].

\bibitem{Chang:2022nzu}
Z.~Chang, Y.-T.~Kuang, X.~Zhang and J.-Z.~Zhou, \emph{{Primordial black holes
  and third order scalar induced gravitational waves*}},
  \href{https://doi.org/10.1088/1674-1137/acc649}{\emph{Chin. Phys. C}
  {\bfseries 47} (2023) 055104}
  [\href{https://arxiv.org/abs/2209.12404}{{\ttfamily 2209.12404}}].

\bibitem{Wang:2023sij}
S.~Wang, Z.-C.~Zhao and Q.-H.~Zhu, \emph{{Constraints On Scalar-Induced
  Gravitational Waves Up To Third Order From Joint Analysis of BBN, CMB, And
  PTA Data}},  \href{https://arxiv.org/abs/2307.03095}{{\ttfamily 2307.03095}}.

\bibitem{Inomata:2020cck}
K.~Inomata, \emph{{Analytic solutions of scalar perturbations induced by scalar
  perturbations}},
  \href{https://doi.org/10.1088/1475-7516/2021/03/013}{\emph{JCAP} {\bfseries
  03} (2021) 013} [\href{https://arxiv.org/abs/2008.12300}{{\ttfamily
  2008.12300}}].

\bibitem{Chang:2022dhh}
Z.~Chang, X.~Zhang and J.-Z.~Zhou, \emph{{The cosmological vector modes from a
  monochromatic primordial power spectrum}},
  \href{https://doi.org/10.1088/1475-7516/2022/10/084}{\emph{JCAP} {\bfseries
  10} (2022) 084} [\href{https://arxiv.org/abs/2207.01231}{{\ttfamily
  2207.01231}}].

\bibitem{Chang:2020tji}
Z.~Chang, S.~Wang and Q.-H.~Zhu, \emph{{Note on gauge invariance of second
  order cosmological perturbations}},
  \href{https://doi.org/10.1088/1674-1137/ac0c74}{\emph{Chin. Phys. C}
  {\bfseries 45} (2021) 095101}
  [\href{https://arxiv.org/abs/2009.11025}{{\ttfamily 2009.11025}}].

\bibitem{Wang:2019kaf}
S.~Wang, T.~Terada and K.~Kohri, \emph{{Prospective constraints on the
  primordial black hole abundance from the stochastic gravitational-wave
  backgrounds produced by coalescing events and curvature perturbations}},
  \href{https://doi.org/10.1103/PhysRevD.99.103531}{\emph{Phys. Rev. D}
  {\bfseries 99} (2019) 103531}
  [\href{https://arxiv.org/abs/1903.05924}{{\ttfamily 1903.05924}}].

\bibitem{Planck:2018vyg}
{\scshape Planck} collaboration, \emph{{Planck 2018 results. VI. Cosmological
  parameters}},
  \href{https://doi.org/10.1051/0004-6361/201833910}{\emph{Astron. Astrophys.}
  {\bfseries 641} (2020) A6}
  [\href{https://arxiv.org/abs/1807.06209}{{\ttfamily 1807.06209}}].

\bibitem{Saikawa:2018rcs}
K.~Saikawa and S.~Shirai, \emph{{Primordial gravitational waves, precisely: The
  role of thermodynamics in the Standard Model}},
  \href{https://doi.org/10.1088/1475-7516/2018/05/035}{\emph{JCAP} {\bfseries
  05} (2018) 035} [\href{https://arxiv.org/abs/1803.01038}{{\ttfamily
  1803.01038}}].

\bibitem{Lepage:1977sw}
G.P.~Lepage, \emph{{A New Algorithm for Adaptive Multidimensional
  Integration}}, \href{https://doi.org/10.1016/0021-9991(78)90004-9}{\emph{J.
  Comput. Phys.} {\bfseries 27} (1978) 192}.

\bibitem{Lepage:2020tgj}
G.P.~Lepage, \emph{{Adaptive multidimensional integration: VEGAS enhanced}},
  \href{https://doi.org/10.1016/j.jcp.2021.110386}{\emph{J. Comput. Phys.}
  {\bfseries 439} (2021) 110386}
  [\href{https://arxiv.org/abs/2009.05112}{{\ttfamily 2009.05112}}].

\bibitem{peter_lepage_2023_10402580}
P.~Lepage, \emph{gplepage/vegas: vegas version 5.6},  Dec., 2023.
\newblock 10.5281/zenodo.10402580.

\bibitem{Siemens:2013zla}
X.~Siemens, J.~Ellis, F.~Jenet and J.D.~Romano, \emph{{The stochastic
  background: scaling laws and time to detection for pulsar timing arrays}},
  \href{https://doi.org/10.1088/0264-9381/30/22/224015}{\emph{Class. Quant.
  Grav.} {\bfseries 30} (2013) 224015}
  [\href{https://arxiv.org/abs/1305.3196}{{\ttfamily 1305.3196}}].

\bibitem{Robson:2018ifk}
T.~Robson, N.J.~Cornish and C.~Liu, \emph{{The construction and use of LISA
  sensitivity curves}},
  \href{https://doi.org/10.1088/1361-6382/ab1101}{\emph{Class. Quant. Grav.}
  {\bfseries 36} (2019) 105011}
  [\href{https://arxiv.org/abs/1803.01944}{{\ttfamily 1803.01944}}].

\bibitem{Zhao:2022kvz}
Z.-C.~Zhao and S.~Wang, \emph{{Bayesian Implications for the Primordial Black
  Holes from NANOGrav\textquoteright{}s Pulsar-Timing Data Using the
  Scalar-Induced Gravitational Waves}},
  \href{https://doi.org/10.3390/universe9040157}{\emph{Universe} {\bfseries 9}
  (2023) 157} [\href{https://arxiv.org/abs/2211.09450}{{\ttfamily
  2211.09450}}].

\bibitem{lamb2023rapid}
W.G.~Lamb, S.R.~Taylor and R.~van Haasteren, \emph{Rapid refitting techniques
  for bayesian spectral characterization of the gravitational wave background
  using pulsar timing arrays}, {\emph{Physical Review D} {\bfseries 108} (2023)
  103019}.

\bibitem{mitridate2023ptarcade}
A.~Mitridate, D.~Wright, R.~von Eckardstein, T.~Schröder, J.~Nay, K.~Olum
  et~al., \emph{Ptarcade},  2023.

\bibitem{Gorji:2023sil}
M.A.~Gorji, M.~Sasaki and T.~Suyama, \emph{{Extra-tensor-induced origin for the
  PTA signal: No primordial black hole production}},
  \href{https://doi.org/10.1016/j.physletb.2023.138214}{\emph{Phys. Lett. B}
  {\bfseries 846} (2023) 138214}
  [\href{https://arxiv.org/abs/2307.13109}{{\ttfamily 2307.13109}}].

\bibitem{Chang:2022vlv}
Z.~Chang, X.~Zhang and J.-Z.~Zhou, \emph{{Gravitational waves from primordial
  scalar and tensor perturbations}},
  \href{https://doi.org/10.1103/PhysRevD.107.063510}{\emph{Phys. Rev. D}
  {\bfseries 107} (2023) 063510}
  [\href{https://arxiv.org/abs/2209.07693}{{\ttfamily 2209.07693}}].

\bibitem{Chen:2022dah}
C.~Chen, A.~Ota, H.-Y.~Zhu and Y.~Zhu, \emph{{Missing one-loop contributions in
  secondary gravitational waves}},
  \href{https://doi.org/10.1103/PhysRevD.107.083518}{\emph{Phys. Rev. D}
  {\bfseries 107} (2023) 083518}
  [\href{https://arxiv.org/abs/2210.17176}{{\ttfamily 2210.17176}}].

\end{thebibliography}\endgroup

\end{document}